# Time-resolved three-dimensional elucidation of complex-refractive-index alteration induced by ultrashort laser pulses

Running title: 3D Mapping of Filament-Induced Index Change

Takumi Koike*, Yusuke Ito, Naohiko Sugita

*Department of Mechanical Engineering, School of Engineering, The University of Tokyo, Bunkyo-ku, Tokyo 113-8656, Japan*

*Corresponding author: t.koike@mfg.t.u-tokyo.ac.jp (Takumi Koike)*

## Abstract

Ultrashort pulse lasers (USPLs) have garnered attention as tools that can cause various unique phenomena by instantaneously inducing a region of altered physical properties (filaments) in a material. However, a comprehensive understanding of USPL-induced filaments remains elusive owing to the complexity of the dynamics involved and the lack of imaging technology to accurately extract such dynamics. In this study, we propose a novel methodology for measuring the transient properties, i.e., the complex refractive index, of a filament by analyzing the polarization state of a probing pulse. To our best knowledge, our proposed methodology is the first to accurately extract the three-dimensional distribution of the complex refractive index near a filament, which fluctuates ultrafast. Our study provides insights into the complex ablation mechanisms caused by USPLs, which are critical for selecting the optimal laser conditions for micro/nanoprocessing. The findings of this study contribute significantly to condensed matter and computational physics through precise data pertaining to the physical properties of USPL-irradiated regions.

## Introduction

Transparent insulators are essential components in precision electronics, medical devices, and other sophisticated equipment[1–3]. Ultrashort pulse lasers (USPLs) have garnered interest as tools for processing these materials with high precision[4, 5]. One of the reasons that USPLs enable such processing is that their irradiation instantly induces metallic regions (filaments) in the material, thereby converting the irradiated material to an easily machinable state. The behavior of such filaments has been widely investigated in various fields, including precision engineering and condensed matter/computational physics, since their discovery in the late 20th century[6–10]. Therefore, devising an accurate measurement technique for the complex refractive indices of filaments may allow one to address industrial and academic questions surrounding filament behavior.

However, existing techniques are limited in their ability to fulfill this purpose. For example, shadowgraphy, which is a method for measuring the imaginary component of the refractive index[11], cannot measure the real component, and its measurement results depend significantly on the intensity of the observation light. Interferometry, which is a technique for measuring the real component[12], requires meticulous experimental techniques to induce optical interference at the observation point and is thus susceptible to errors in the measurement results. Researchers have attempted to capture all components of the refractive index, such as via pseudorecovery by combining the abovementioned two (or equivalent) methods; however, the effectiveness remains limited owing to challenges in capturing exactly the same moment using different methods. Ellipsometry can effectively capture all components of the complex refractive index[13]. However, it is not applicable to filaments since analysis by this method basically relies on the probing light reflected from the material surface, which prevents the analysis of the internal properties. While several studies have used transmitted probes for ellipsometry, to our knowledge, no study has used this method to analyze microstructures such as filaments.

In this study, we propose a novel method that can accurately measure a filament by analyzing the polarization state of an observation light transmitted through an object. To our best knowledge, this is the first approach that can precisely extract the spatiotemporal evolution of all components of the complex refractive index of a filament. The results show reasonable agreement with both conventional measurements and numerical calculations, thereby confirming the accuracy of the present approach.

## Results

### Concept of Proposed Method

The experimental configuration shown in Fig. 1 is designed to analyze the polarization state of the observation pulse transmitted through a filament. A single USPL emitted from the amplifier in the setup separates into two branches. One branch (processing pulse) is used to generate a filament in the sample, whereas the other branch (observation pulse) is employed to observe high-speed phenomena induced by the former. Temporal separation of the two pulses is established via a motor-controlled optical delay to perform a time-resolved measurement. At a fixed optical delay, a series of images is obtained while the polarizer angle is varied systematically from 0° to 180° (see Appendix 1 for details). This operation enables the measurement of the light-intensity behavior as a function of the angle display, $\alpha$, of the rotating polarizer, at a fixed time delay after the start of filament formation. Hereinafter, the distribution of this experimentally obtained light-intensity behavior is denoted as $\iota(\alpha)$ (see Fig. 2). The details of the experimental conditions are presented in Table 1.

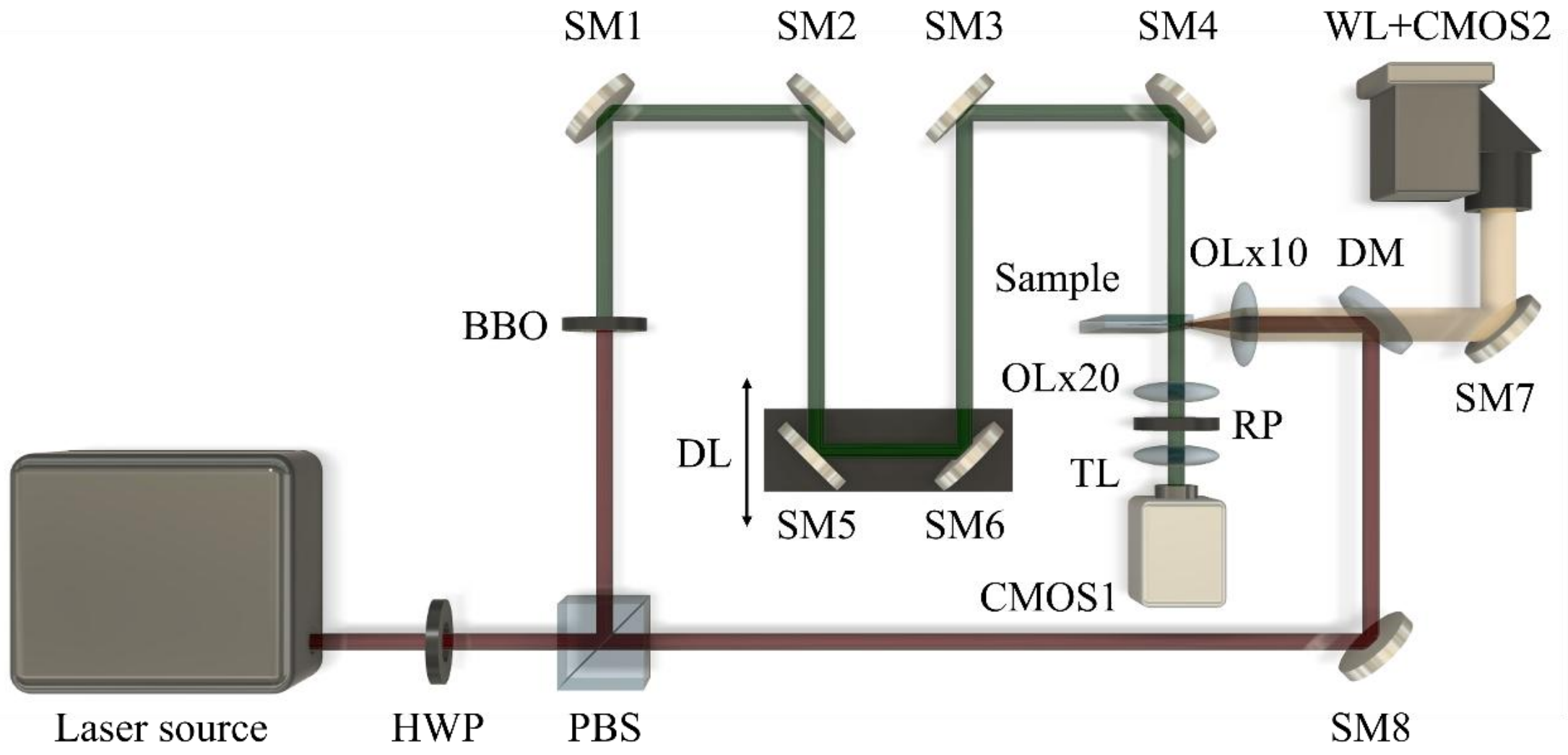

**Fig. 1 Schematic illustration of experimental setup.** HWP: half-wave plate; PBS: polarizing beam splitter; BBO: beta barium borate crystal; SM: silver mirror; DL: motor-controlled optical delay line; DM: dichroic mirror; OLx10/x20: objective lens; TL: tube lens; CMOS1: BH-73M, Bitran; CMOS2: DCC1645C-HQ, Thorlabs; WL: white light. WL is utilized for sample positioning: sample position is precisely adjusted such that its surface is imaged onto CMOS2, which corresponds to the surface located at the focal plane of OLx10. This allows precise determination of USPL focusing position within the material.

**Table 1 Experimental conditions**

| | |
|---|---|
| Sample | Silica glass |
| Pulse energy | 100 μJ |
| Pulse duration | 2 ps |
| Focal position | 100 μm under surface |

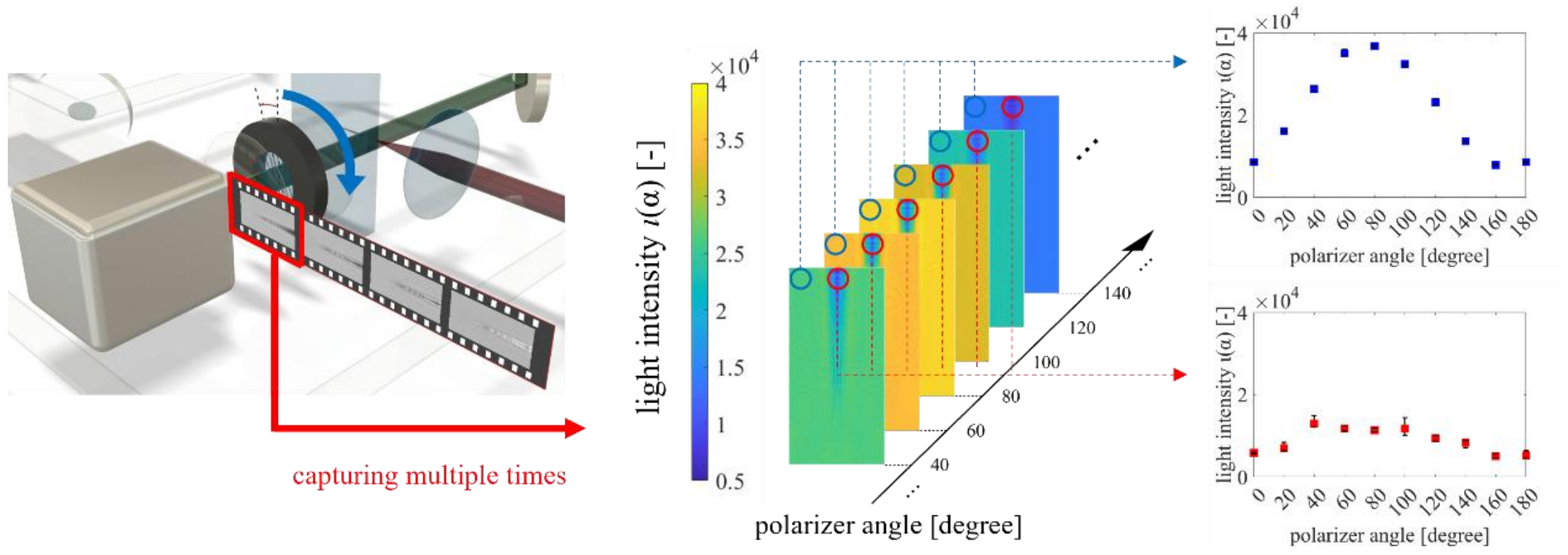

**Fig. 2 Schematic illustration of light-intensity measurement method.** At a fixed optical delay, filament images are recorded while rotating the polarizer positioned in front of the imaging device. This enables the measurement of the intensity profile, $\iota(\alpha)$, as a function of the polarizer angle at each pixel.

The behavior of the recorded light intensity, $\iota(\alpha)$, can theoretically be expressed by $I(\alpha)$ (see Eq. (1)). This expression is derived by squaring the absolute value of the electric field, $\boldsymbol{E}(\alpha)$, which is obtained using the Jones calculus (Eq. (2), see Appendix 2 for a concise derivation). In this equation, $\theta$ denotes the angle subtended by the transmission axis of the rotating polarizer with the vertical axis when the angle display is at the origin. The remaining parameters ($\Psi$, $\Delta$, $\psi$, and $\delta$) are employed to describe the polarization state of the observation pulse. The first two parameters describe the polarization state change after propagation through the filament, whereas the last two parameters describe the initial polarization state. The values of these parameters can be estimated by identifying the representation of $I(\alpha)$ that yields the closest fit to $\iota(\alpha)$ using the least-squares-approximation method (see Appendix 3 for details).

$$
\begin{aligned}
I(\alpha)
&= A|\boldsymbol{E}(\alpha)|^2 \\
&= A|-\sin\Psi\sin\psi\exp(-j(\Delta+\delta))\sin(\alpha+\theta)+\cos\Psi\cos\psi\cos(\alpha+\theta)|^2,
\end{aligned}
\tag{1}
$$

$$
\begin{aligned}
\boldsymbol{E}(\alpha)
&\propto \begin{bmatrix} 0 & 0 \\ 0 & 1 \end{bmatrix}\begin{bmatrix} \cos(\alpha+\theta) & \sin(\alpha+\theta) \\ -\sin(\alpha+\theta) & \cos(\alpha+\theta) \end{bmatrix}\begin{bmatrix} \sin\Psi\exp(-j\Delta) & 0 \\ 0 & \cos\Psi \end{bmatrix} \\
&\quad\times\begin{bmatrix} \sin\psi\exp(-j\delta) \\ \cos\psi \end{bmatrix} \\
&= \begin{bmatrix} 0 \\ -\sin\Psi\sin\psi\exp(-j(\Delta+\delta))\sin(\alpha+\theta)+\cos\Psi\cos\psi\cos(\alpha+\theta) \end{bmatrix}.
\end{aligned}
\tag{2}
$$

Polarization parameters $\Psi$ and $\Delta$, which are employed to express the amplitude ratio and phase difference of each polarization component of the transmitted observation light[14], are expressed in Eq. (3). The changes in these parameters are induced by the complex-refractive-index distribution near the filament. Using this relation, the complex refractive index can be determined from the polarization parameters, whose values are determined in advance.

$$\tan\Psi \exp(-j\Delta) \equiv \frac{t_P}{t_S}. \tag{3}$$

Fig. 3 illustrates this relationship schematically. Fig. 3(a) presents the geometric relationship between a filament and the observation pulse, where the radial and axial directions are denoted by the *r*- and *z*-axes, respectively. This representation is based on a reasonable assumption that the filament is axially symmetric, since the intensity profile of the USPL that induces the filament follows a Gaussian distribution. The primary factor contributing to this polarization-state change is the inhomogeneity of the refractive index. The contributing phenomena include Fresnel effects, multiple internal reflections at refractive index boundaries, and birefringence[14]. Among these, the effects of Fresnel refractions and multiple internal reflections are negligible in absorptive media[15] or in media with gradually varying refractive indices (see Appendix 4). Therefore, birefringence was considered the dominant contributor in the present analysis.

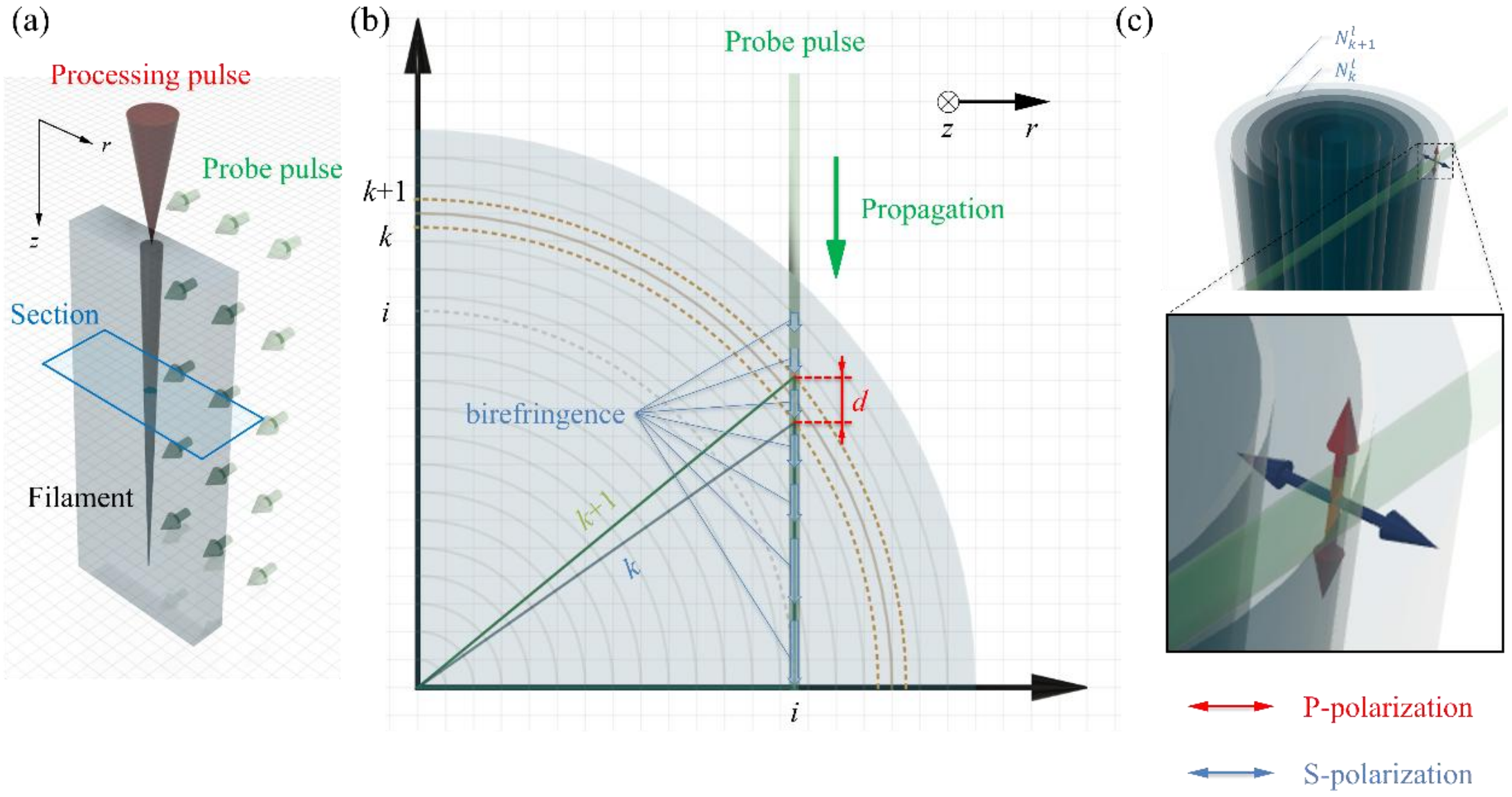

**Fig. 3 Relationship between USPL-induced filament and change in polarization state of probe pulse.** (a) Perspective view, (b) Cross-sectional view at $z = l$, and (c) refractive-index distribution for each polarization direction.

The effect of birefringence can be expressed as follows[16–18]:

$$\begin{aligned}&\boldsymbol{\delta}\frac{t_P}{t_S}\bigg|_{r=k\sim k+1}^{z=l}\\&=\boldsymbol{\delta}\{\tan\Psi\exp(-j\Delta)\}|_{r=k\sim k+1}^{z=l}\\&=\exp\left\{\frac{2\pi jd}{\lambda}(n_P-n_S)\right\}.\end{aligned} \tag{4}$$

This equation describes the complex-transmittance-ratio change (and thus the polarization-state change; see Eq. (3)) induced in the observation pulse that intersects at $r = i$ as it propagates through the concentric refractive-index layers from $r = k + 1$ to $r = k$ (see Fig. 3(b)). Here, $d$ denotes the spacing between the two concentric layers; $\lambda$ is the wavelength; $n_P$ and $n_S$ are the refractive indices for the P- and S-polarized components, respectively; and $j$ is an imaginary unit. The symbol $\boldsymbol{\delta}$ signifies a minor alteration in the physical quantity and is unrelated to the polarization parameter $\delta$. Under these configurations, the S- and P-polarization directions are oriented horizontally and in-plane, respectively (see Fig. 3(b)). Accordingly, $n_S$ and $n_P$ can be identified with $N_{k+1}^l$ and $N_k^l$, respectively (see Fig. 3(c)), and Eq. (4) can be reformulated as Eq. (5), where $N_k^l$ denotes the refractive index at coordinates $(r, z) = (k, l)$.

$$\exp\left\{\frac{2\pi jd}{\lambda}(n_P-n_S)\right\}=\exp\left\{\frac{2\pi jd_k^i}{\lambda}\left(N_k^l-N_{k+1}^l\right)\right\},$$

where

$$d_k^i\equiv\begin{cases}\sigma\left(\sqrt{(k+1)^2-i^2}-\sqrt{k^2-i^2}\right)\cdots\text{for } k\geq i+1\\ \sigma\sqrt{2i+1}\qquad\cdots\text{for } k=i\end{cases}. \tag{5}$$

Here, $\sigma$ denotes the pixel length in the image and $d_k^i$ is the effective spacing between the concentric layers as viewed from the observation coordinate $r = i$, which can be determined through simple geometric considerations.

The cumulative change in the polarization state after passing through all the concentric layers can be expressed as a product of the individual contributions at each position. This is expressed in Eq. (6), where $i_{max}$ is the $r$-axis endpoint coordinate of the imaging area.

$$\tan\Psi\exp(-j\Delta)|_{r=i}^{z=l} = \prod_{k=i}^{i_{max}}\left[\exp\left\{\frac{2\pi d_k^i j}{\lambda}\left(N_k^l - N_{k+1}^l\right)\right\}\right]^2. \quad (6)$$

Utilizing Eq. (6), the complex refractive index near the filament can be obtained. A detailed explanation of this procedure is provided in Appendix 5.

**Spatiotemporal Evolution of Polarization Parameters**

Figs. 4(a)–(c) present the experimental results of the polarization parameters in addition to the corresponding filaments induced by 2-ps USPLs. Figs. 4(a) and 4(b) show the spatial distributions of polarization parameters $\Psi$ and $\Delta$, respectively, whereas Fig. 4(c) shows the associated filament images obtained using shadowgraphy. A common trend across these results is that at locations distant from the filament region, the polarization parameters approximate the baseline values of ($\Psi$, $\Delta$) = (45°, 0°), which are hereinafter referred to as the ground-state values. By contrast, pronounced deviations from these values are observed near the filaments. This indicates that the proposed method captures spatial regions where the polarization state deviates from the ground state, with a distribution that closely resembles that of the filament.

To enable a more quantitative evaluation, Figs. 5(a)–(c) show the temporal evolution of the polarization parameters and transmittance along the central axis of the same filaments. Each plot is color coded based on the elapsed time following USPL irradiation, where red, magenta, green, and blue represent 2, 10, 100, and 500 ps, respectively. A cross-comparison of the data with identical color coding across the figures reveals the following correlation: a greater deviation of the polarization parameters from their ground-state values is associated with a lower transmittance. Furthermore, the intrafigure temporal progression (i.e., color-coded data within each graph) reveals that both the transmittance and polarization parameters gradually return to their ground-state values as time progresses after irradiation (Figs. 5(a)–(c)). These findings are consistent with the interpretation that USPL-induced refractive index inhomogeneities perturb the polarization state[13] and that the subsequent relaxation of these inhomogeneities enables recovery toward the initial polarization state.

In summary, the present method demonstrates high sensitivity to polarization changes arising from refractive-index modulations near USPL-induced filaments, thereby providing a robust method for characterizing ultrafast laser–material interactions.

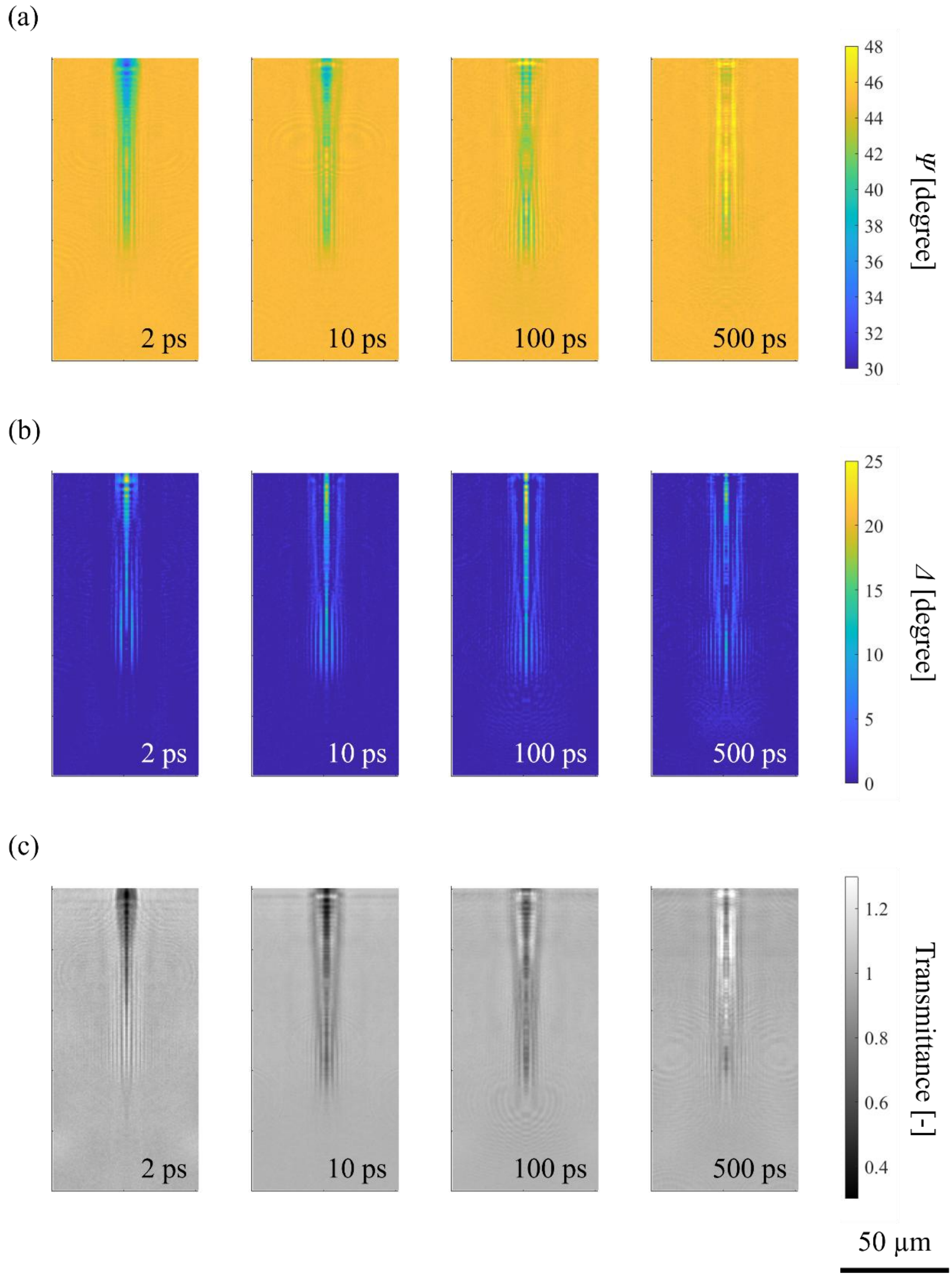

**Fig. 4 Spatiotemporal evolution of polarization parameters and corresponding filaments.** (a) $\Psi$ [°], (b) $\Delta$ [°], and (c) transmittance [-].

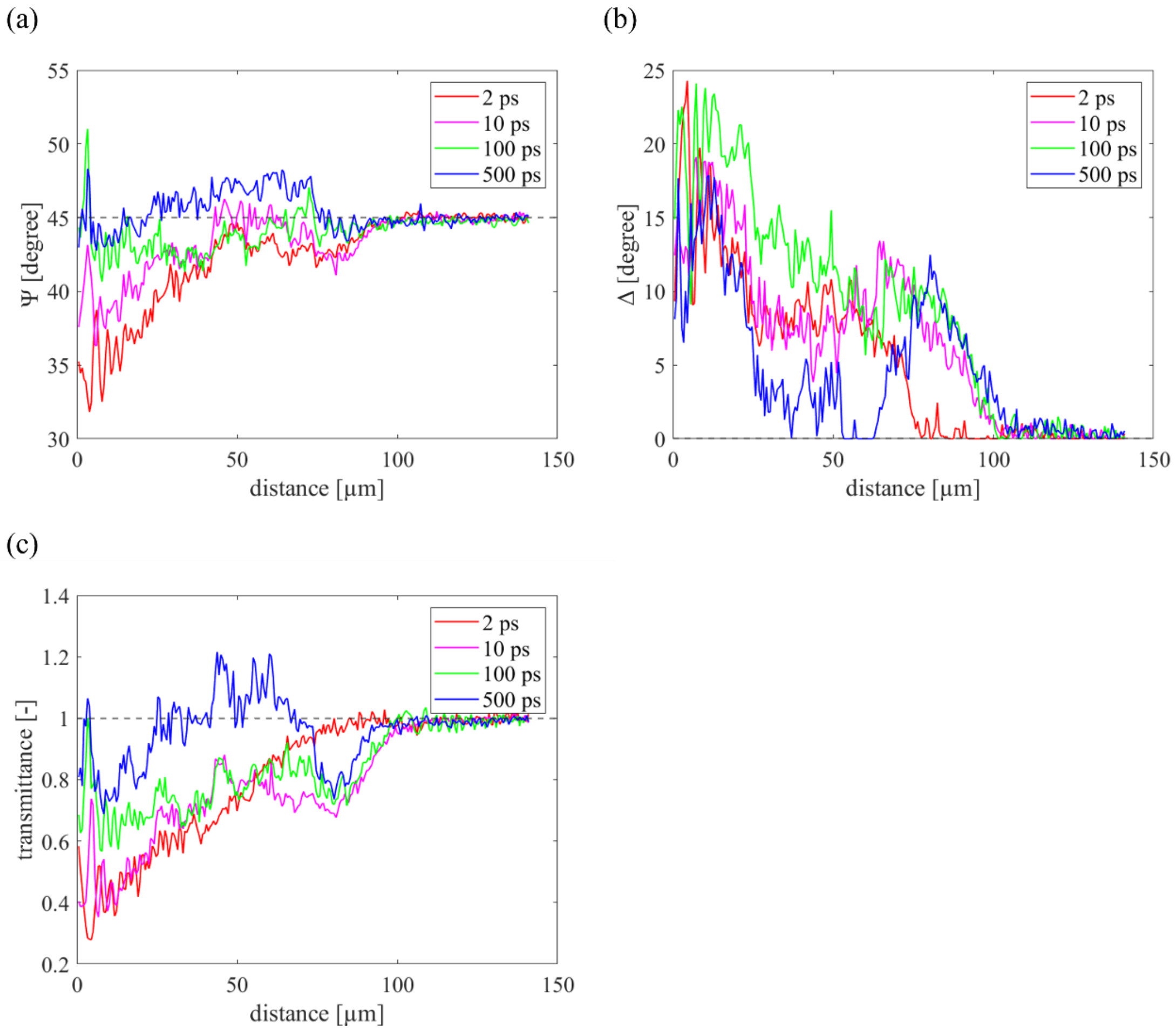

**Fig. 5 Temporal evolution of polarization parameters and transmittance along central axis of filaments.** (a) *Ψ* [°], (b) *Δ* [°], and (c) transmittance [-]. Plots are color coded based on elapsed time after USPL irradiation: red, magenta, green, and blue correspond to 2, 10, 100, and 500 ps, respectively.

## Refractive-Index Distribution in 2-ps USPL-Induced Filaments

Fig. 6 shows the spatial distribution of the complex refractive index of the filaments obtained using the proposed method. The regions where the refractive index deviates from that of pristine fused silica ($n_0$ = 1.461) are almost identical to the locations where the filaments are induced (see Fig. 4(c)).

In general, the deviation of the refractive index from $n_0$ is governed by the density of photoexcited electrons within the material. Regions subjected to stronger laser fields, where more significant electron excitation occurs, tend to exhibit larger deviations. These regions are typically aligned along the central axis of the filament, and the radial profile of the refractive-index change is expected to exhibit a Gaussian distribution, thereby reflecting the spatial intensity profile of the incident laser field. Furthermore, as time progresses, the excited electrons gradually relax, thus resulting in a corresponding reduction in the refractive-index deviation. The present experimental results show good qualitative agreement with these general characteristics.

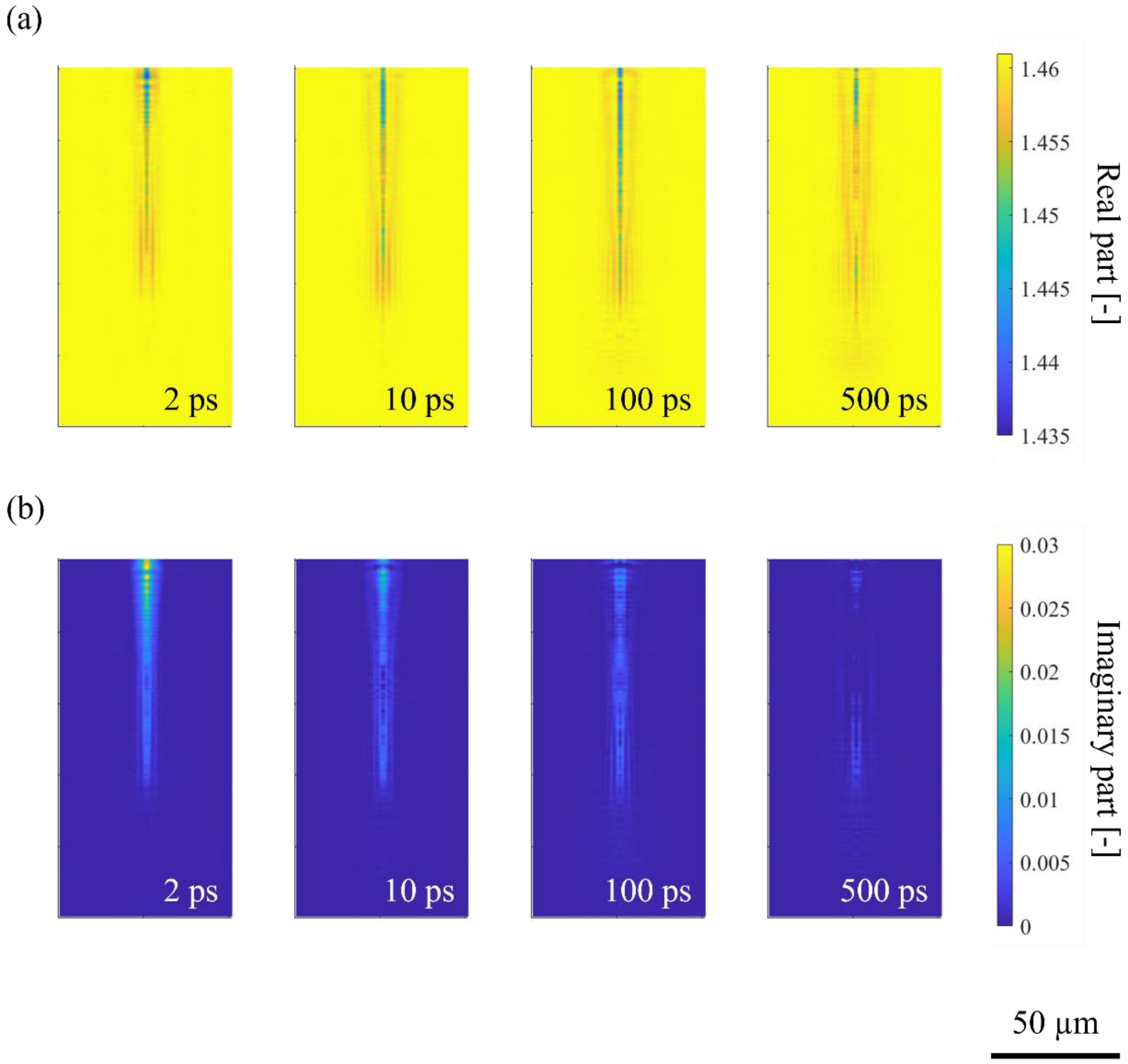

**Fig. 6 Spatiotemporal evolution of complex refractive index of filament obtained using proposed method.** (a) real component [-], and (b) imaginary component [-].

### Validation of Proposed Method via Comparative Analysis Using Shadowgraphy

The reliability of the results obtained using the proposed method was assessed via comparative analysis using a conventional experimental technique and numerical simulations (for details, see Appendices 6 and 7, respectively). First, the conventional method was employed to evaluate the imaginary component of the refractive index. Specifically, shadowgraphy was utilized, which leverages the established correlation between the transmittance of the probe beam and the material's absorption coefficient. The results obtained using this approach, as presented in Fig. 7(a), exhibit high qualitative consistency with those obtained using the proposed method (Fig. 6(b)). In both cases, the deviation of the refractive index from $n_0$ was spatially localized along the filament axis, with radial profiles exhibiting Gaussian-like attenuation. Furthermore, the temporal evolution of the measured signals was consistent with the generally accepted relaxation dynamics of photoexcited electrons in the materials, which was likewise consistent with the results obtained using the proposed method. These findings provide indirect but convincing evidence that supports the reliability of the proposed method.

To enable a quantitative comparison, Fig. 7(b) shows the axial profiles obtained using both the proposed technique and shadowgraphy. For each delay time following USPL irradiation, the data acquired using both methods were superimposed, with those acquired using the proposed method and shadowgraphy shown in red and blue plots, respectively. This comparison reveals a high degree of quantitative agreement between the two approaches. For instance, at a delay time of 2 ps, both methods yielded values ranging from 0.025 to 0.020 over a propagation distance of 0–25 μm. Furthermore, both profiles exhibited similar asymptotic decays toward zero with almost identical curvatures, as shown in Fig. 7(b1). This level of agreement was indicated consistently at all subsequent delay times, as illustrated in Figs. 7(b2)–(b4). These results indicate that the measurements obtained using the proposed method are in excellent quantitative agreement with those derived from the conventional technique under all tested conditions. Consequently, the accuracy of the proposed method in evaluating the imaginary component is considered comparable to or potentially superior to that of the established measurement techniques (see Appendix 6 for a more comprehensive discussion).

**Numerical Validation Based on Drude Model**

To further evaluate the reliability of the proposed method, numerical simulations were conducted. In these simulations, the real component of the refractive index corresponding to the experimentally obtained imaginary component was computationally estimated and compared with the measured values. This approach was justified by the established relationship between the real and imaginary components of the refractive index, which can be reasonably inferred using a classical model describing excited electrons in dielectrics (the Drude model). As detailed in Appendix. 7, this relationship is governed by the electron collision frequency, $v$, which is a parameter characterizing the dynamics of excited carriers. Reported values of $v$ are typically in the range of $1 \times 10^{15}$–$1 \times 10^{16}$ $s^{-1}$ (refs. 19–24), which constrains the plausible values of the real component. Consequently, good agreement between the computed and measured real components, using a $v$ value within this reported range, may be regarded as indirect evidence supporting the validity of the proposed method.

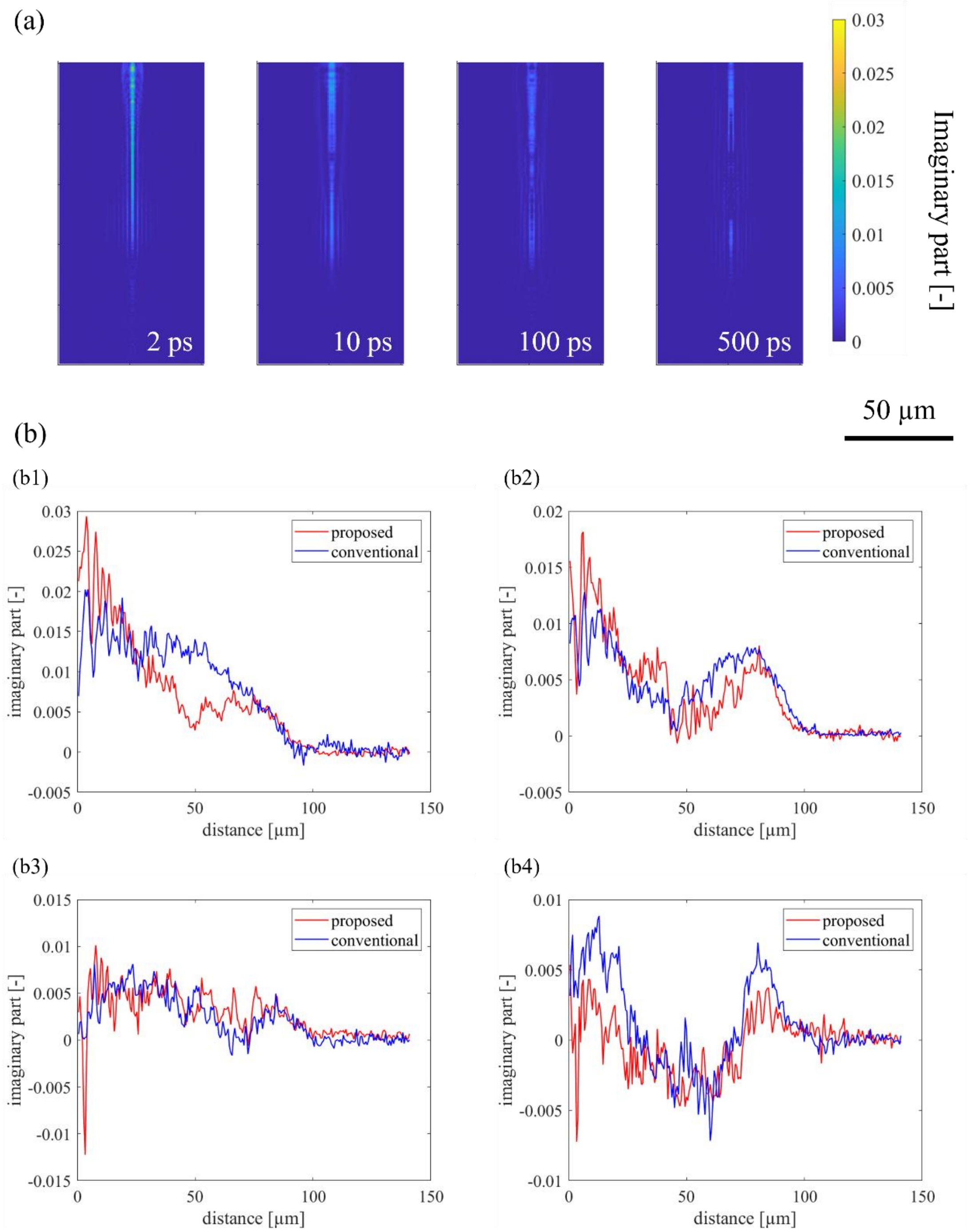

**Fig. 7. Validation of proposed measurement method using shadowgraphy.** (a) Spatiotemporal evolution of imaginary component measured using shadowgraphy. (b) Comparison of measured results along central axis of filament; red and blue plots represent results obtained using proposed method and shadowgraphy, respectively.

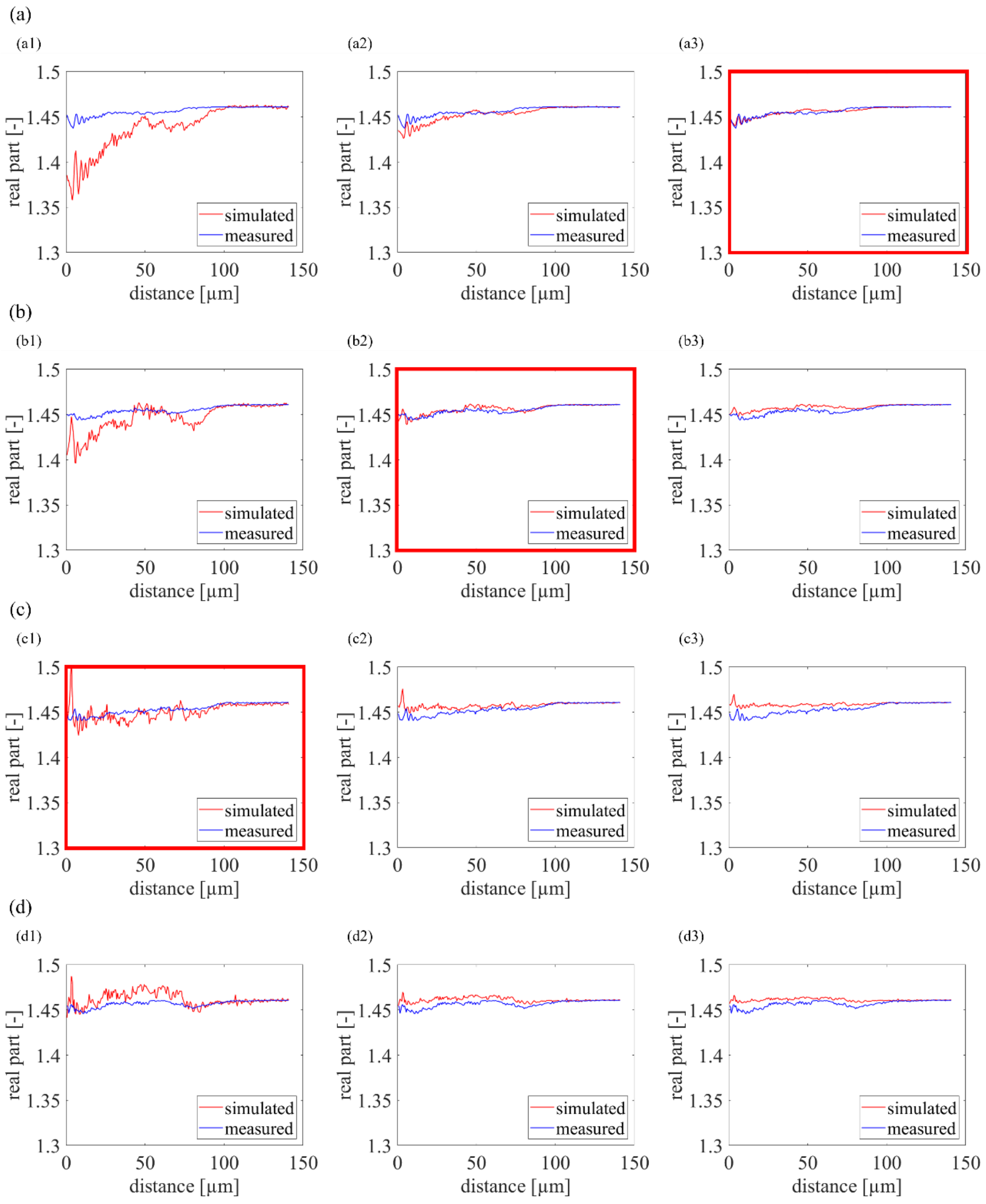

**Fig. 8. Validation of proposed measurement method through comparison with simulation results.** Blue plots represent results obtained using proposed method while red plots indicate corresponding simulation results. Panels (a)–(d) show results at delay times of 2, 10, 100, and 500 ps, respectively. Each panel comprises three subpanels, labeled (x1), (x2), and (x3), which present simulation results obtained with electron collision frequencies of $\nu = 1\times10^{15}$, $3\times10^{15}$, and $5\times10^{15}$ $s^{-1}$, respectively. Subpanel showing closest agreement between experimental and simulated results is enclosed by red frame.

Fig. 8 presents the results of the numerical simulations. Panels (a)–(d) show overlays of the experimentally measured values (blue curves) and the corresponding numerical estimates (red curves), both of which were obtained at delay times ranging from 2 to 500 ps following USPL irradiation. In the simulations, the electron collision frequency, $v$, was varied across three representative values: $1 \times 10^{15}$, $3 \times 10^{15}$, and $5 \times 10^{15}$ $s^{-1}$.

In the temporal range up to 100 ps post-irradiation (Figs. 8(a)–(c)), the simulations showed good agreement with the experimental data under all the tested conditions. Specifically, the best match at 2 ps was obtained when $v = 5 \times 10^{15}$ $s^{-1}$ (Fig. 8(a3)); at 10 ps, when $v = 3 \times 10^{15}$ $s^{-1}$ (Fig. 8(b2)); and at 100 ps, when $v = 1 \times 10^{15}$ $s^{-1}$ (Fig. 8(c1)). Notably, the collision frequency $v$ that yielded the best agreement between the experimental and numerical results decreased as the delay time increased. This trend is physically consistent with the generally understood relaxation dynamics of excited electrons: $v$ reflects the rate of interactions between excited electrons and other carriers or phonons and is typically higher during the early stages of excitation when electronic activity is the most intense[19–24]. Accordingly, the observed temporal decrease in $v$ can be interpreted as indicative of progressive electron relaxation and a concomitant reduction in their kinetic energy.

By contrast, at a delay of 500 ps (Fig. 8(d)), a clear discrepancy was observed between the simulation and experimental results. A plausible explanation for this deviation is that, at this stage, the material state exceeds the validity range of the Drude model. Indeed, optical luminescence—phenomena not observed at earlier times—is detected near the filament under these conditions (Figs. 4(c) and 5(c)), suggesting the involvement of additional processes such as plasma luminescence that lie beyond the descriptive scope of the Drude model. Accordingly, the disagreement observed at this later time is attributed to the limitations of the underlying model, rather than a flaw in the proposed method itself. Nevertheless, a more comprehensive understanding of phenomena occurring on this timescale requires further investigation.

## Discussion

This study presents a novel method for spatiotemporally resolving alterations in the complex refractive index induced by USPL irradiation in transparent dielectrics. Whereas USPLs are known to generate transient, filamentary structures characterized by localized changes in material properties, the precise measurement of these dynamics—particularly that of the complex refractive index—remains challenging owing to the limitations of existing diagnostic techniques. Hence, we propose a polarization-based imaging technique that analyzes the transmitted probe beam's polarization state to extract the full complex-refractive-index distribution in three dimensions with femtosecond temporal resolution. Notably, the proposed method enables the simultaneous detection of the real and imaginary components of the refractive index, thereby overcoming a major limitation of conventional techniques such as interferometry and shadowgraphy, which typically isolate only a single component. The present method successfully captured the evolution of refractive-index variations in fused silica, thus revealing a strong correlation between refractive-index modulation and the expected temporal-relaxation behavior. The accuracy of the results was validated through quantitative comparisons with conventional shadowgraphy and further supported by numerical simulations based on the Drude model. Good agreement was observed over a range of delay times up to 100 ps, with the electron collision frequency decreasing over time, which is in accordance with the expected relaxation dynamics of excited carriers. At longer timescales (> 500 ps), deviations from the simulation results were attributed to the breakdown of the Drude model, which is possibly due to the emerging plasma luminescence, thus suggesting a complex regime warranting further investigation.

In conclusion, this study introduced a robust and generalizable technique for visualizing and quantifying ultrafast light–matter interactions, with significant implications for optimizing USPL-based micro/nanofabrication and advancing the understanding of non-equilibrium processes in dielectric media.

## Materials and methods

An ultrashort laser (Pharos SP-10W, Light Conversion) with a wavelength of 1030 nm and a pulse duration of 2 ps was employed as the light source. The laser beam was divided into two branches: a processing pulse and an observation pulse. The processing pulse was focused 100 μm below the surface of a silica glass sample (AGC, AQ; dimensions: 0.2 × 25 × 75 $mm^3$, all surfaces polished) using an objective lens (Mitutoyo, M Plan Apo NIR 10×), thereby inducing a plasma filament within the material. The pulse energy of the processing beam was set to 100 μJ at the sample position. The observation pulse was frequency-doubled using a beta barium borate (BBO) crystal (15-278, Edmund Optics) and transmitted through the filament region to probe its polarization-dependent transmittance. Temporal separation between the processing and observation pulses was controlled by a motorized optical delay line (OSMS26-300 (X), OptoSigma), allowing time-resolved measurements at four delay times: 2 ps, 10 ps, 100 ps, and 500 ps, respectively. At each fixed delay, a series of images was recorded while systematically varying the angle of a rotating polarizer from 0° to 180°. This procedure enabled measurement of the transmitted intensity as a function of the polarizer angle (see Appendix 1 for details). The filament region was magnified using an additional objective lens (Mitutoyo, M Plan Apo NIR 20×), which served to enlarge the observed area and improve the spatial resolution of the measurements. The imaging system comprised two CMOS cameras: the BH-73M (Bitran) recorded the observation beam transmitted through the filament, while the DCC1645C-HQ (Thorlabs) monitored the sample surface and facilitated alignment. For alignment, white light illuminated the sample surface via the OL×10 objective lens. The reflected white light from the surface was imaged onto the DCC1645C-HQ camera. Under this configuration, the sample surface appeared sharpest when positioned at the OL×10 lens' focal plane. This image was used to precisely determine the sample position and thereby accurately align the laser focus within the material.

## Acknowledgements

We express our gratitude to the Japan Society for the Promotion of Science (24KJ0748), JST PRESTO (JPMJPR22Q1), and MEXT Quantum Leap Flagship Program (JPMXS0118067246).

## Author contributions

Conceptualization: T.K. Methodology: T.K. Investigation: T.K. Visualization: T.K. Funding acquisition: N.S., Y.I., and T.K. Supervision: N.S. and Y.I. Writing—original draft: T.K. Writing—review and editing: T.K and Y.I.

## Conflict of interest

The authors have no conflicts to disclose.

## Data availability

Data underlying the results presented herein are not publicly available at this time but may be obtained from the authors upon reasonable request.

# Supplementary Material for

## Time-resolved and three-dimensional elucidation of complex-refractive-index alteration induced by ultrashort laser pulses

Takumi Koike[1*], Yusuke Ito[1], Naohiko Sugita[1]

*[1]Department of Mechanical Engineering, School of Engineering, The University of Tokyo, 7-3-1 Hongo, Bunkyo, Tokyo 113-8656, Japan*

* *Corresponding author: t.koike@mfg.t.u-tokyo.ac.jp (Takumi Koike)*

## 1. Details on Data-Processing Methodology

Under a fixed optical delay between the processing and observation pulses, the angle of the polarizer in the setup was systematically varied in 20° increments from 0° to 180°, which resulted in 10 distinct measurement conditions (see the left diagram in Fig. 2). For each polarizer angle condition, two types of images were acquired: one captured without USPL irradiation (background (BG) images) and the other captured during filament induction (filament images). Images were acquired on three separate occasions, which yielded three pairs of filament and the corresponding BG images for each condition. All images were saved in raster format; each pixel presents a real-valued intensity corresponding to the detected light at that position, with higher values indicating stronger intensity. The following preprocessing steps were applied to the data:

I. Each filament image was normalized by dividing its pixel values by those of the corresponding BG image, thus resulting in intensity ratios typically ranging from 0 to 1.
II. The normalized image was multiplied by the average pixel value of the corresponding BG image. This correction step mitigates fluctuations in the overall light intensity arising from factors unrelated to material properties, such as variations in the laser amplifier or imaging system.
III. The corrected images obtained under each polarizer angle condition were averaged across the three measurement sessions to reduce inter-experimental variability.
IV. Assuming axial symmetry of the refractive-index distribution, symmetric images were generated by averaging the pixel values on both sides of the filament axis. In subsequent analyses, only the axisymmetric half of the image was retained.

By performing these procedures, the light-intensity distribution as a function of polarizer angle, denoted $\iota(\alpha)$, was obtained for each condition (see the right diagram in Fig. 2).

## 2. Jones Calculus-Based Derivation of Eq. (2)

In this section, the derivation of Eq. (2) is presented based on the Jones calculus, the fundamentals of which are described in (ref. 14). For clarity and conciseness, the notations summarized in Table A1 are employed throughout this derivation. Using these abbreviations, Eq. (2) can be rewritten as

$$\boldsymbol{E}(\alpha) \propto \boldsymbol{P}_y \boldsymbol{M}_R(\alpha + \theta)\boldsymbol{S}\boldsymbol{J}_{in}. \tag{A1}$$

**Table A1 Abbreviations and corresponding Jones matrices/vectors.**

| Optical elements | Abbreviations | Jones matrix/vector |
|---|---|---|
| Polarizer | $\boldsymbol{P}_y$ | $\begin{bmatrix} 0 & 0 \\ 0 & 1 \end{bmatrix}$ |
| Rotation matrix | $\boldsymbol{M}_{\mathrm{R}}(\varphi)$ | $\begin{bmatrix} \cos\varphi & \sin\varphi \\ -\sin\varphi & \cos\varphi \end{bmatrix}$ |
| Transmission of material | $\boldsymbol{S}$ | $\begin{bmatrix} \sin\Psi\exp(-j\Delta) & 0 \\ 0 & \cos\Psi \end{bmatrix}$ |
| Incident electric field | $\boldsymbol{J}_{\mathrm{in}}$ | $\begin{bmatrix} \sin\psi\exp(-j\delta) \\ \cos\psi \end{bmatrix}$ |

Multiplication of the incident Jones vector $\boldsymbol{J}_{\mathrm{in}}$ by the matrix $\boldsymbol{S}$ from the left yields the polarization state after transmission through the material. This is based on the relationship between the transmittance ratio of the P- and S-polarized components $t_{\mathrm{P}}/t_{\mathrm{S}}$ and the ellipsometric parameters $\Psi$ and $\Delta$.

$$\tan\Psi\exp(-j\Delta) \equiv \frac{t_P}{t_S} = \frac{E'_P/E_P}{E'_S/E_S} \Longrightarrow \frac{E'_P}{E'_S} = \tan\Psi\exp(-j\Delta)\frac{E_P}{E_S}. \tag{A2}$$

Thus, the transmitted electric field vector can be written as

$$\therefore \begin{bmatrix} E'_P \\ E'_S \end{bmatrix} \propto \begin{bmatrix} \sin\Psi\exp(-j\Delta) & 0 \\ 0 & \cos\Psi \end{bmatrix}\begin{bmatrix} E_P \\ E_S \end{bmatrix} = \boldsymbol{S}\boldsymbol{J}_{in}. \tag{A3}$$

Here, $E_{\mathrm{P/S}}$ and $E'_{\mathrm{P/S}}$ denote the electric field amplitudes for the P-/S-components before and after transmission, respectively.

Subsequent multiplication by the rotation matrix $\boldsymbol{M}_{\mathrm{R}}(\alpha + \theta)$ and the polarizer matrix $\boldsymbol{P}_y$ in this order yields the electric field after passing through the analyzer, whose transmission axis is tilted by an angle $\alpha + \theta$ from the vertical. This process is illustrated schematically in Fig. A1. Specifically, $\boldsymbol{M}_{\mathrm{R}}(\alpha + \theta)$ represents a counterclockwise rotation of the coordinate system by $\alpha + \theta$ (Fig. A1(b)) and $\boldsymbol{P}_y$ corresponds to a polarizer whose transmission axis is aligned with the new $y'$-axis (Fig. A1(c)).

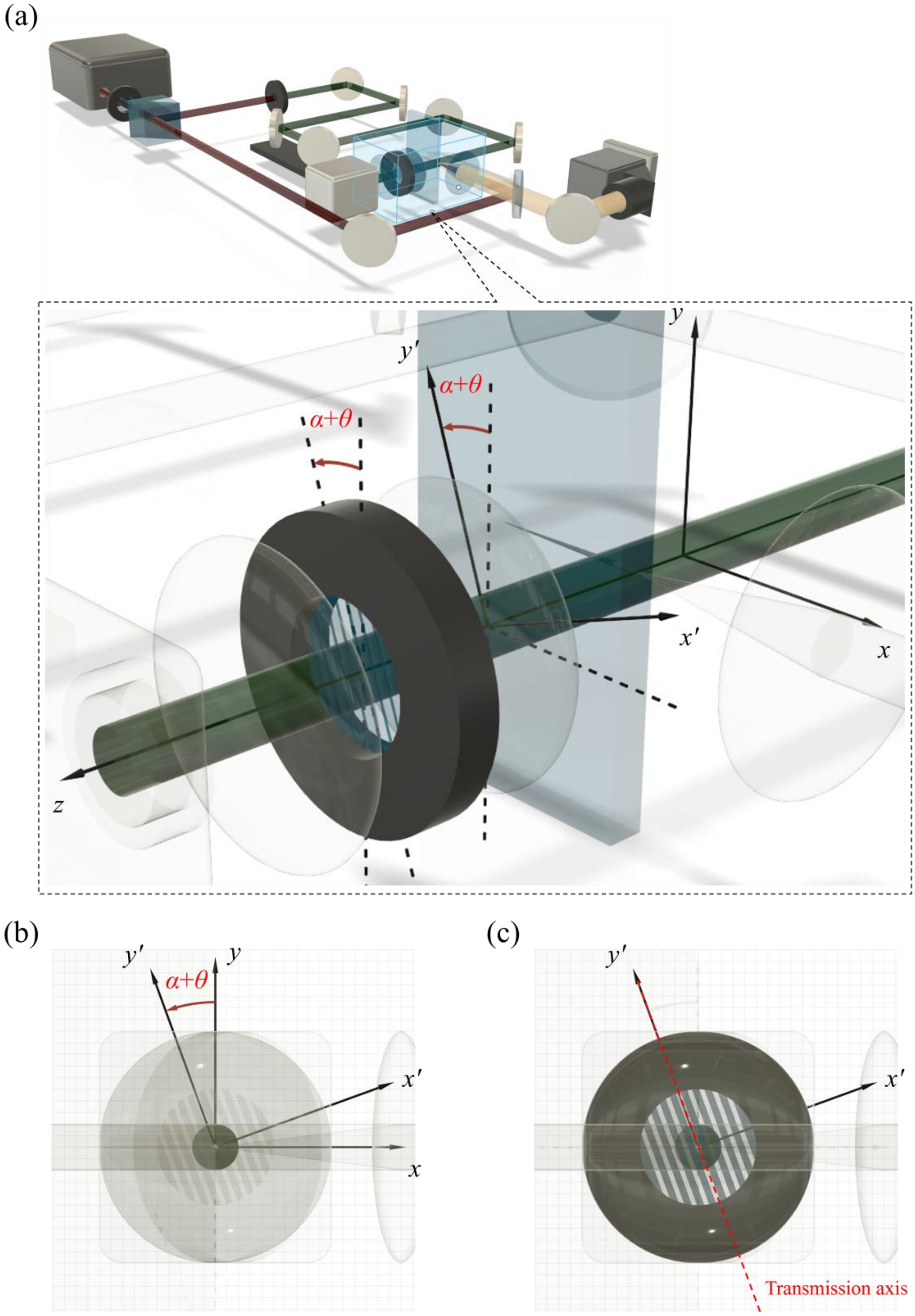

**Fig. A1 Schematic representation of Jones-matrix-based modeling of light transmission and analysis.** (a) Overall optical setup corresponding to Eq. (A1); (b) definition of rotated coordinate system by angle $\alpha + \theta$; (c) orientation of polarizer along $y'$-axis.

## 3. Estimation Procedure for Polarization Parameters $\Psi$ and $\Delta$

Using the least-squares-approximation method, the polarization parameters $\Psi$ and $\Delta$ were estimated by identifying the theoretical expression $I(\alpha)$ that best fits the experimentally measured light intensity $\imath(\alpha)$. This estimation process involves multiple steps, as several additional parameters—the initial orientation angle of the polarizer $\theta$ and the initial polarization parameters of the observation pulse $\psi$ and $\delta$—must be determined. The following subsections provide a detailed description of each step involved in this estimation procedure.

### 3-1. Determination of Initial Orientation Angle $\theta$

The orientation angle $\theta$ of the polarizer cannot be known a priori owing to the mechanical ambiguity introduced by the engagement of threaded components in the polarizer assembly, specifically that between the mounted wire-grid polarizer (WP25M-UB1, Thorlabs) and the continuous rotation mount (RSP1D/M, Thorlabs). To determine $\theta$, the experimentally measured intensity $\imath'(\alpha)$ for an S-polarized pulse—whose polarization state is known in advance—was compared with the theoretical intensity $I'(\alpha)$, which is expressed in Eq. (A5). The value of $\theta$ was obtained by minimizing the least-squares error between $\imath'(\alpha)$ and $I'(\alpha)$.

$$\begin{aligned}\boldsymbol{E}'(\alpha) &\propto \begin{bmatrix}0 & 0\\ 0 & 1\end{bmatrix}\begin{bmatrix}\cos(\alpha+\theta) & \sin(\alpha+\theta)\\ -\sin(\alpha+\theta) & \cos(\alpha+\theta)\end{bmatrix}\begin{bmatrix}0\\ 1\end{bmatrix}\\ &= \begin{bmatrix}0\\ \cos(\alpha+\theta)\end{bmatrix},\end{aligned} \tag{A4}$$

$$\begin{aligned}I'(\alpha) &= A'|\boldsymbol{E}'(\alpha)|^2\\ &= A'\cos^2(\alpha+\theta).\end{aligned} \tag{A5}$$

Here, the measured intensity $\imath'(\alpha)$ was spatially averaged over the entire imaging region based on the assumption that light propagating in free space ideally exhibits a uniform intensity distribution. The spatial fluctuations in the measured data were therefore treated as noise and excluded from the estimation of $\theta$. Subsequently, the value of $\theta$ obtained in this step was applied in all the following estimations.

### 3-2. Estimation of Initial Polarization Parameters $\psi$ and $\delta$

The initial polarization parameters $\psi$ and $\delta$ were introduced to account for changes induced by second-harmonic generation (SHG) in the nonlinear optical crystal. Since the polarization state after SHG varies depending on the crystal properties, these parameters cannot be assumed a priori. To estimate these values, the experimentally measured intensity $\imath''(\alpha)$ was compared with its theoretical expression $I''(\alpha)$. The optimal parameters were again obtained by minimizing the least-squares error.

$$
\begin{aligned}
&\boldsymbol{E}''(\alpha) \\
&\propto \begin{bmatrix} 0 & 0 \\ 0 & 1 \end{bmatrix} \begin{bmatrix} \cos(\alpha+\theta) & \sin(\alpha+\theta) \\ -\sin(\alpha+\theta) & \cos(\alpha+\theta) \end{bmatrix} \begin{bmatrix} \sin\psi \exp(-j\delta) \\ \cos\psi \end{bmatrix} \\
&= \begin{bmatrix} 0 \\ -\sin\psi \exp(-j\delta) \sin(\alpha+\theta) + \cos\psi \cos(\alpha+\theta) \end{bmatrix},
\end{aligned}
\tag{A6}
$$

$$
\begin{aligned}
&I''(\alpha) \\
&= A''|\boldsymbol{E}''(\alpha)|^2 \\
&= A''|-\sin\psi \exp(-j\delta) \sin(\alpha+\theta) + \cos\psi \cos(\alpha+\theta)|^2.
\end{aligned}
\tag{A7}
$$

The waveform $I''(\alpha)$ may appear almost flat either when the amplitude $A''$ is small or when specific values of $\psi$ and $\delta$ result in destructive interference (see Fig. A2). To remove this ambiguity, the mean intensity over all polarizer angles was evaluated, which resulted in the relation:

$$
\frac{1}{\pi}\int_0^{\pi} A''|-\sin\psi \exp(-j\delta) \sin(\alpha+\theta) + \cos\psi \cos(\alpha+\theta)|^2 d\alpha = \frac{A''}{2}. \tag{A8}
$$

The equation above allows $A''$ to be determined independently from the mean of $\imath''(\alpha)$, thus enabling a unique estimation of $\psi$ and $\delta$ through least-squares fitting.

In practice, this estimation can be integrated with the procedure described in Section 3-3. Specifically, the measured intensity $\imath''(\alpha)$ may be replaced by the spatially averaged intensity over regions sufficiently distant from the filament, where the polarization state after SHG is assumed to remain unaltered. This assumption is justified by the experimental geometry: the probe light enters perpendicularly to the side surface of the sample, such that polarization modifications other than SHG are expected to be localized near the filament.

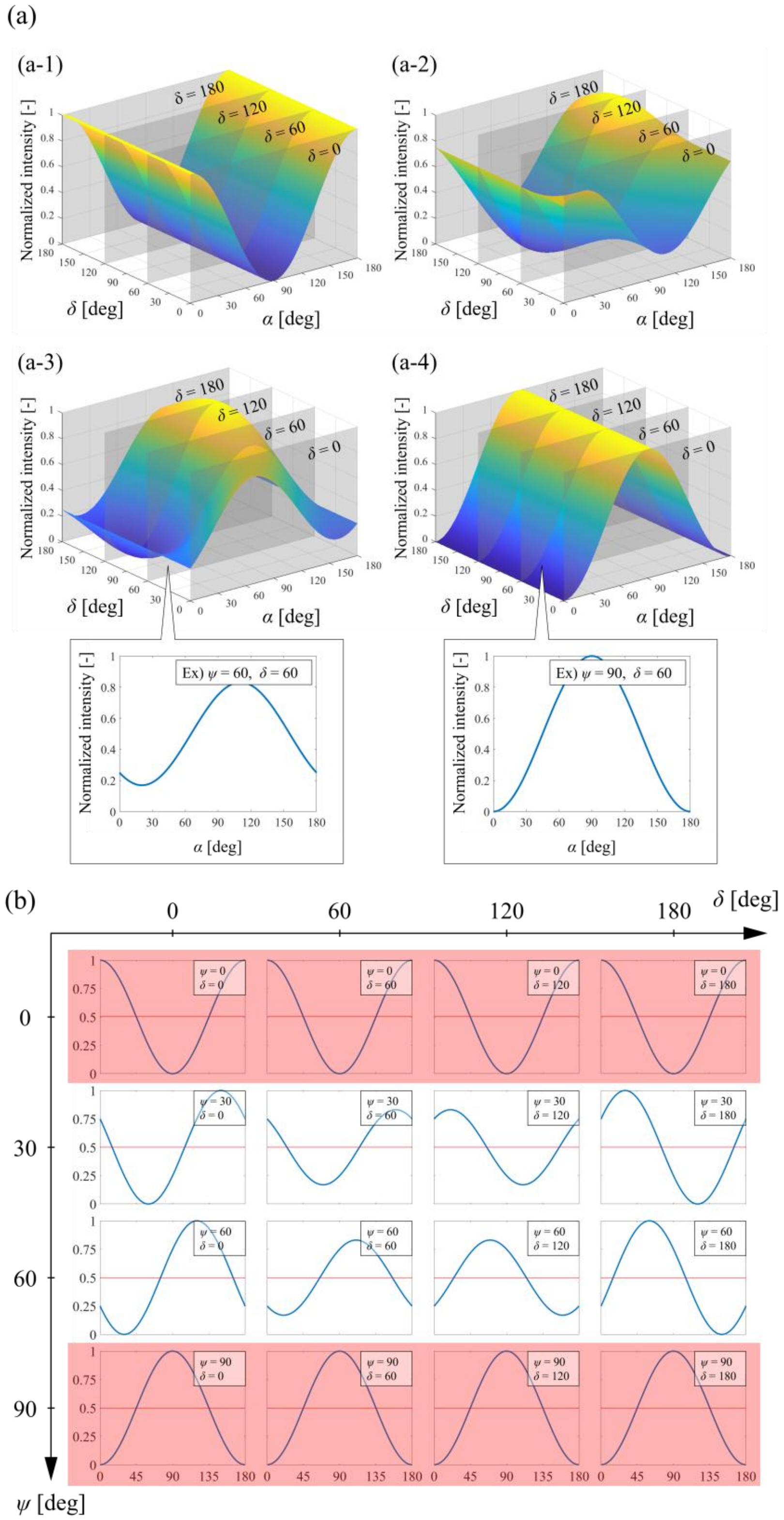

**Fig. A2 Dependence of normalized theoretical intensity $I''(\alpha)$ on polarization parameters $\psi$, $\delta$, and $\alpha$.** (a) Representative surface plots; (a-1)–(a-4) surface views for fixed $\psi$ (= 0°, 30°, 60°, and 90°) and varying $\delta$ and $\alpha$; (b) tiled cross-sectional profiles of $I''(\alpha)$ vs. $\alpha$ for various combinations of $\psi$ and $\delta$. Red-shaded regions in panel (b) indicate parameter domains with potential least-squares fitting instability, which are avoided in this study.

### 3-3. Estimation of Polarization Parameters *Ψ* and *Δ*

Finally, the previously determined parameters $\theta$, $\psi$, and $\delta$ were used to estimate the polarization parameters of interest, $\Psi$ and $\Delta$, by least-squares fitting the theoretical intensity waveform $I(\alpha)$ to the measured data $\iota(\alpha)$.

Fig. A3 presents the representative normalized waveforms as functions of $\Psi$ and $\Delta$. Unlike in the previous step, these waveforms do not exhibit distinctive features beyond the fact that their mean values are determined solely by $\Psi$. Consequently, the analytical strategy employed in Section 3-2 cannot be directly applied here. Nonetheless, the waveform can be characterized by three independent features—its mean intensity, the deviation of its extrema from the mean, and its phase offset—corresponding to three unknowns: the amplitude $A$ and the polarization parameters $\Psi$ and $\Delta$. Therefore, simultaneous determination of all three parameters is, in principle, feasible. Whereas the same principle formally applies to the analysis in Section 3-2, the robustness of parameter estimation generally improves when the number of free parameters is minimized. Hence, the amplitude $A''$ was preliminarily estimated prior to the determination of $\psi$ and $\delta$ in the previous subsection.

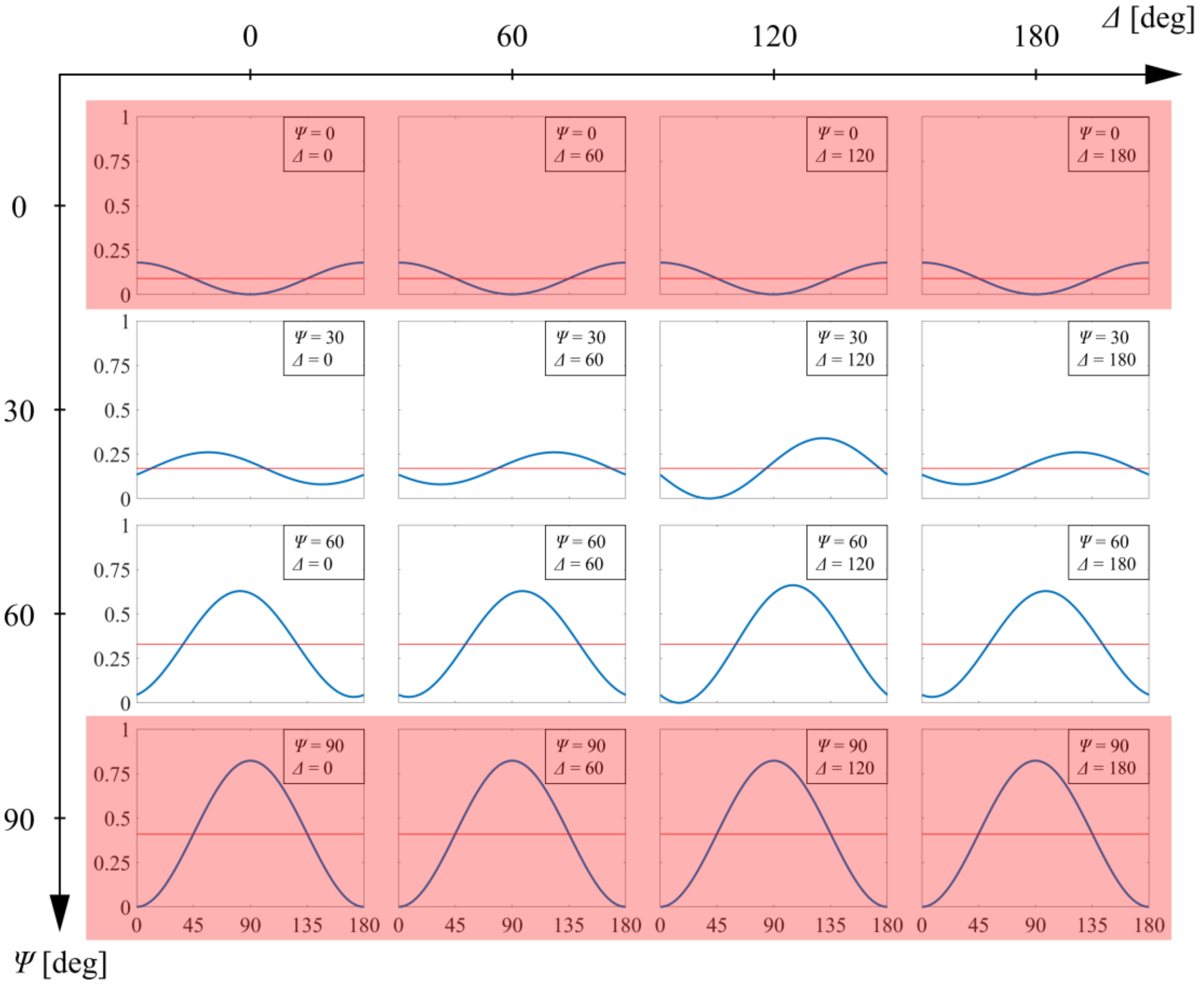

**Fig. A3 Tiled profiles of normalized $I(\alpha)$ as functions of polarization parameters $\Psi$ and $\Delta$**, with red-shaded regions indicating potential least-squares fitting instability, which are avoided in this study.

## 4. Analysis of Fresnel Effects in Polarization

In the main text, birefringence was assumed to be the dominant mechanism underlying the observed variations in the polarization state of transmitted light. This supplementary section presents a quantitative evaluation to validate this assumption. As mentioned previously, additional contributions may arise from multiple reflections at interfaces between regions of differing refractive indices as well as from Fresnel refractions at such boundaries. Among these, the effect of multiple reflections is generally negligible in absorbing media such as filaments[15]. Consequently, the present analysis focuses on quantifying the influence of Fresnel effects.

In this regard, an extended analytical framework was employed that incorporated both birefringence and Fresnel effects while maintaining the same optical configuration used in the main text (see Fig. A4). The resulting refractive-index distribution obtained from this extended framework was then compared with that obtained using the birefringence-only model described in the main text. Hereafter in this appendix, results derived from the extended analysis are referred to as the *Fresnel-inclusive results*, whereas those obtained using the birefringence-only model are denoted as the *Reference results*. A close agreement between the two indicates that Fresnel effects contribute marginally to the observed polarization changes.

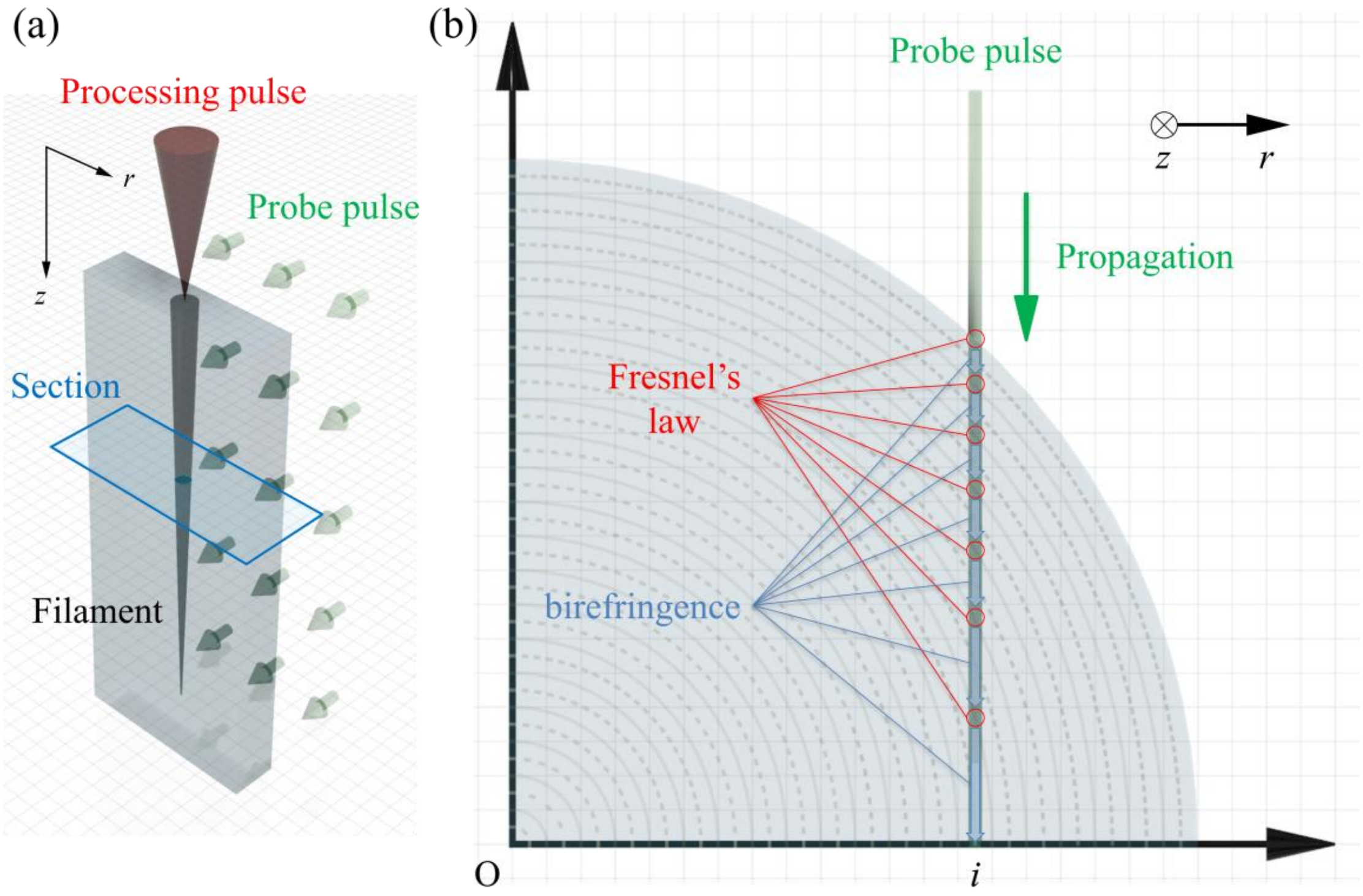

**Fig. A4 Relationship between USPL-induced filament and polarization-state change of probe pulse.** (a) Perspective view; (b) cross-sectional view at $z = l$ incorporating both Fresnel effects and birefringence.

To quantitatively evaluate the contribution of Fresnel effects, we derived Eq. (A9), which describes the change in the transmission ratio and, accordingly (cf. Eq. (3)), the corresponding alteration in the polarization state. This formulation accounts for the interaction of an observation pulse with concentric refractive-index layers from $r = k + 1$ to $r = k$. The angles $\beta$ and $\beta'$ denote the incidence and refraction angles at the interface, respectively (see Fig. A5). $\boldsymbol{\delta}_{\text{Fresnel}}$ signifies a minor alteration in the physical quantity induced by Fresnel effects. All other parameters and coordinates are defined in accordance with those adopted in the main text.

$$\begin{aligned} &\boldsymbol{\delta}_{\text{Fresnel}} \frac{t_P}{t_S}\bigg|_{r=k\sim k+1}^{z=l} \\ &= \boldsymbol{\delta}_{\text{Fresnel}}\{\tan\Psi\exp(-j\Delta)\}|_{r=k\sim k+1}^{z=l} \\ &= \frac{N_{k+1}^{l}\cos\beta + N_k^l\cos\beta\,'}{N_k^l\cos\beta + N_{k+1}^{l}\cos\beta\,'}; \end{aligned} \tag{A9}$$

Here, the angle $\beta'$ in Eq. (A9) satisfies Snell's law.

$$N_k^l\sin\beta\,' = N_{k+1}^l\sin\beta. \tag{A10}$$

Meanwhile, the incident angle $\beta$ is geometrically determined from triangle △OAB in Fig. A5 and is expressed as

$$\beta = \sin^{-1}\left(\frac{i}{k+0.5}\right). \tag{A11}$$

Substituting Eqs. (A10) and (A11) into Eq. (A9) yields

$$\begin{aligned} &\boldsymbol{\delta}_{\text{Fresnel}}\{\tan\Psi\exp(-j\Delta)\}|_{r=k\sim k+1}^{z=l} \\ &= \frac{N_{k+1}^l\cos\left\{\sin^{-1}\left(\frac{i}{k+0.5}\right)\right\} + N_k^l\sqrt{1-\left(\frac{N_{k+1}^l\frac{i}{k+0.5}}{N_k^l}\right)^2}}{N_k^l\cos\left\{\sin^{-1}\left(\frac{i}{k+0.5}\right)\right\} + N_{k+1}^l\sqrt{1-\left(\frac{N_{k+1}^l\frac{i}{k+0.5}}{N_k^l}\right)^2}}. \end{aligned} \tag{A12}$$

The cumulative effect on the polarization state after propagation through all concentric layers—considering both birefringence and Fresnel effects—is expressed as a product of the individual contributions at each segment. This is expressed by combining Eq. (6) from the main text with Eq. (A12), where $i_{\max}$ denotes the terminal radial coordinate within the imaging region.

$$\tan\Psi\exp(-j\Delta)|_{r=i}^{z=l} = \prod_{k=i}^{i_{max}}\left[\boldsymbol{\delta}\{\tan\Psi\exp(-j\Delta)\}|_{r=k\sim k+1}^{z=l}\cdot\boldsymbol{\delta}_{\mathrm{Fresnel}}\{\tan\Psi\exp(-j\Delta)\}|_{r=k\sim k+1}^{z=l}\right]^2. \tag{A13}$$

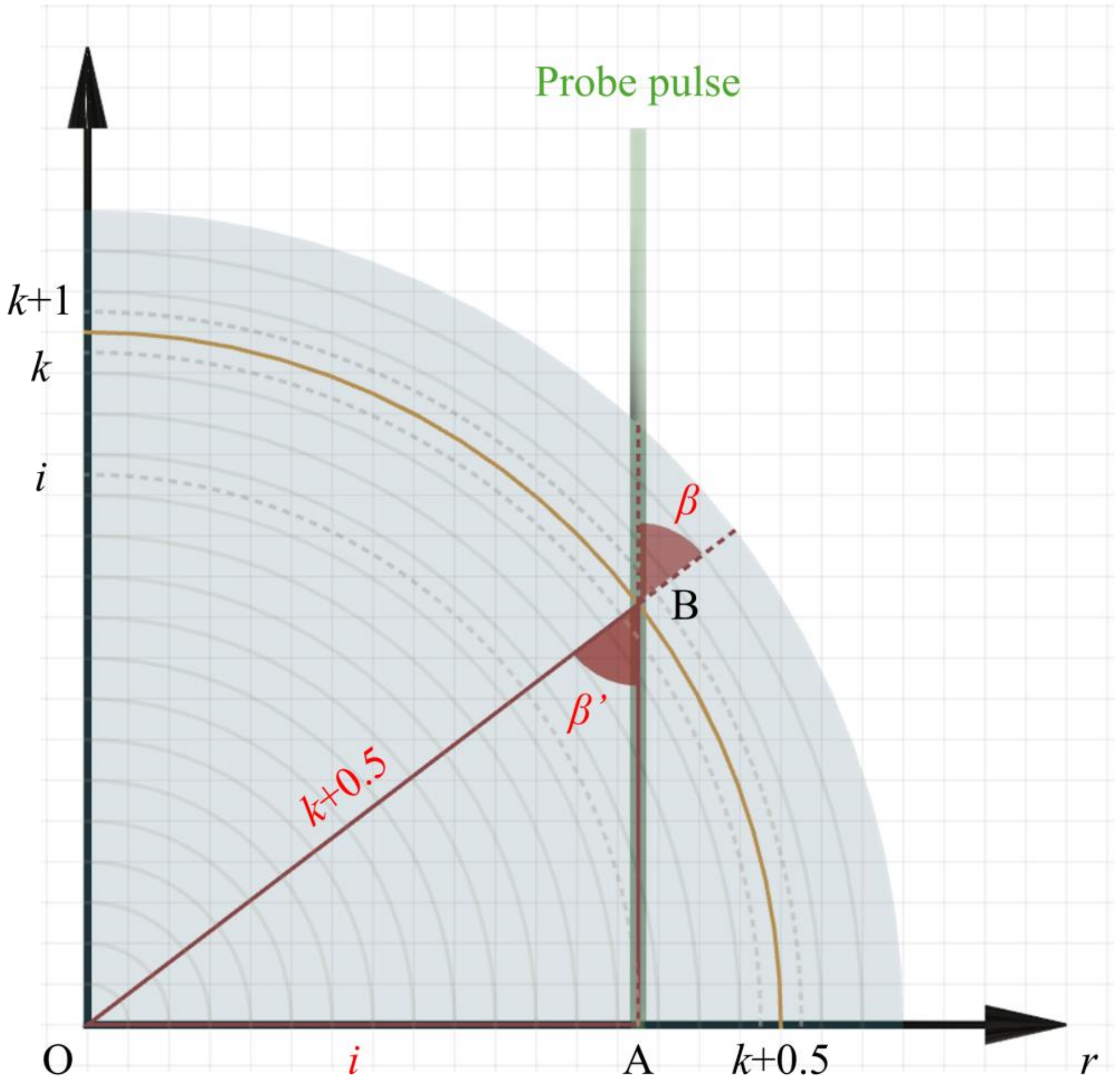

**Fig. A5 Cross-sectional geometry of filament with relevant parameters for quantitative Fresnel-effect evaluation.**

Figure A6 presents the spatiotemporal distribution of Fresnel-inclusive results. Panels (a) and (b) show the real and imaginary components, respectively. These results are highly consistent with the Reference results in both spatial and temporal dimensions (cf. Fig. 6), thereby supporting the assumption that Fresnel contributions are marginal.

To quantitatively assess the discrepancy between these results, the relative error was calculated. Specifically, the absolute difference between the two results at each condition was normalized by the absolute value of the corresponding Reference result, and the outcome was expressed as a percentage. This approach enables a direct quantification of the Fresnel effect's contribution to polarization changes in relative terms. The resulting distribution is shown in panel (c). As evident from the figure, the relative error remained below $2.5\times10^{-4}\%$ in all conditions, with the maximum deviation reaching only $2.3683\times10^{-4}\%$. Notably, the observed discrepancy was predominantly localized near the filament axis. This behavior is attributed, at least partly, to the numerical method employed in solving Eq. (A13). Unlike Eq. (A14), this equation cannot be linearized by any known transformation, thus rendering the numerical technique described in Appendix 5 inapplicable. In this analysis, the refractive-index profile was determined sequentially from the outermost radial position toward the center. Consequently, numerical errors inherently accumulated as the calculation proceeded inward, thus resulting in a concentration of errors near the axis of symmetry. By accounting for this inevitable error accumulation inherent to the computational scheme, one can infer that the actual contribution of Fresnel effects is less significant than that indicated in Fig. A6(c). These findings further corroborate the assumption that Fresnel effects minimally affect the polarization state under the present conditions, thereby affirming the validity of the analytical model described by Eq. (6).

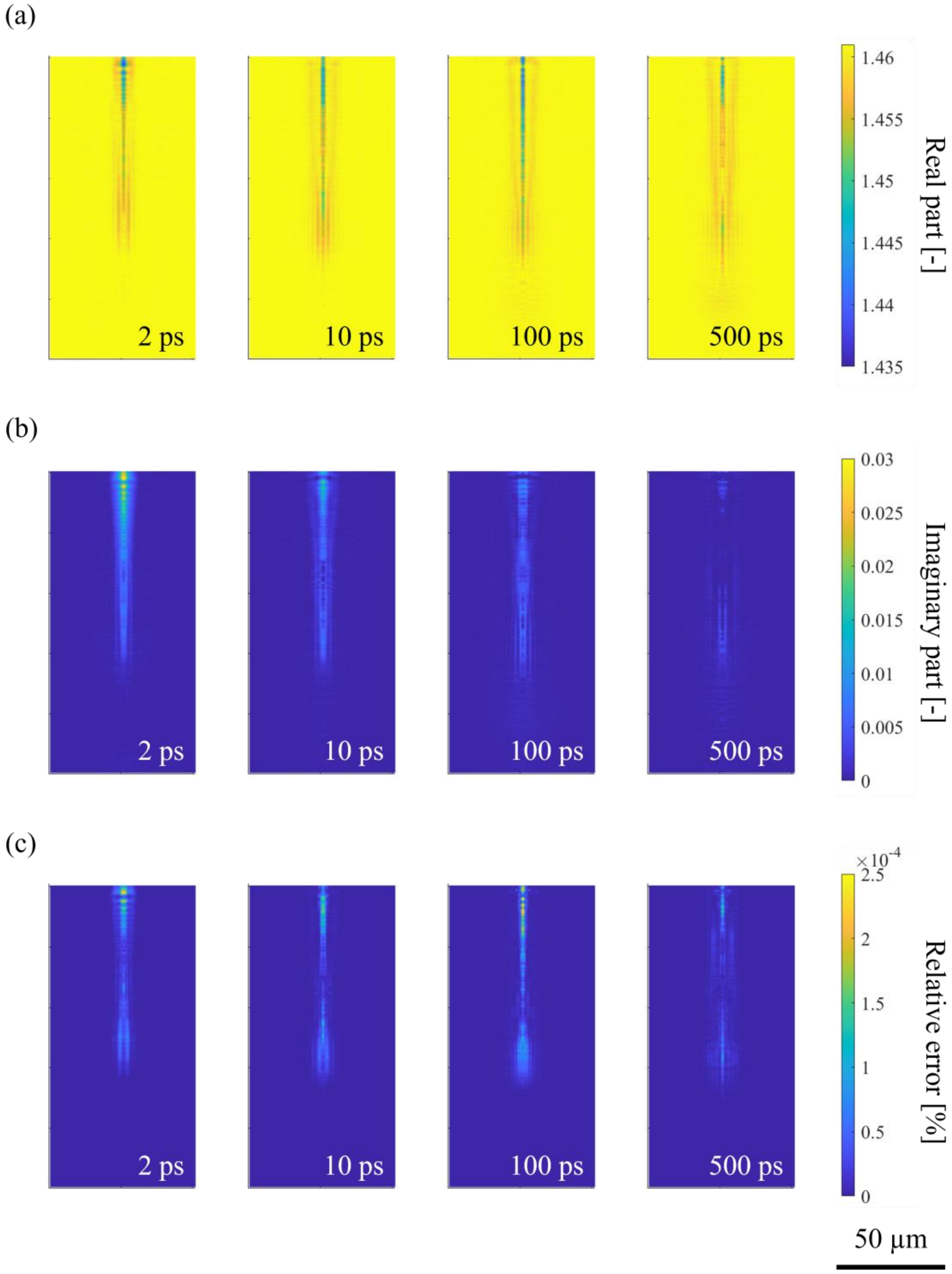

**Fig. A6 Spatiotemporal distribution of Fresnel-inclusive results and relative error compared to Reference results:** (a) Real component; (b) imaginary component of refractive index, and (c) relative error.

## 5. Solution of Eq. (6) Using Matrix Computation

Equation (6) is linearized by applying the natural logarithm to both sides, thus enabling its solution through standard linear algebraic techniques. This transformation allows the simultaneous reconstruction of the refractive index in all radial positions within a specified cross-sectional plane at a fixed axial coordinate *z* while mitigating the accumulation of numerical errors toward the central axis.

Applying the natural logarithm to both sides of Eq. (6) yields the following expression: For notational simplicity, the coordinate designation $z = l$ is omitted in the subsequent discussion. Additionally, the notation $\rho_i \equiv \tan\Psi \exp(-j\Delta)|_{r=i}$ is adopted.

$$\begin{aligned} \rho_i &= \prod_{k=i}^{i_{max}} \left[ \exp\left\{ \frac{2\pi d_k^i j}{\lambda} (N_k - N_{k+1}) \right\} \right]^2 \\ \Rightarrow -\frac{j\lambda}{4\pi} \ln \rho_i &= \sum_{k=i}^{i_{max}} d_k^i \, (N_k - N_{k+1}). \end{aligned} \tag{A14}$$

Equation (A14) can be reformulated as a matrix equation in the following manner. In this formulation, the refractive index at radial positions beyond the imaging region ($r > i_{max}$) is assumed to be equal to $n_0$. This assumption is based on the fact that those regions are sufficiently distant from the area irradiated by the USPL and thus, any refractive-index modulation induced by the irradiation can be regarded as negligible.

$$\begin{aligned} &\frac{-j\lambda}{4\pi} \ln \begin{bmatrix} \rho_1 \\ \rho_2 \\ \vdots \\ \rho_{i_{max}-1} \\ \rho_{i_{max}} \end{bmatrix} \\ &= \begin{bmatrix} a_1 & b_2^1 & \cdots & b_{i_{max}-1}^1 & b_{i_{max}}^1 \\ 0 & a_2 & & b_{i_{max}-1}^2 & b_{i_{max}}^2 \\ \vdots & & \ddots & \vdots & \\ 0 & 0 & \cdots & a_{i_{max}-1} & b_{i_{max}}^{i_{max}-1} \\ 0 & 0 & & 0 & a_{i_{max}} \end{bmatrix} \begin{bmatrix} N_1 \\ N_2 \\ \vdots \\ N_{i_{max}-1} \\ N_{i_{max}} \end{bmatrix} - n_0 \begin{bmatrix} d_{i_{max}}^1 \\ d_{i_{max}}^2 \\ \vdots \\ d_{i_{max}}^{i_{max}-1} \\ d_{i_{max}}^{i_{max}} \end{bmatrix} \end{aligned} \tag{A15}$$

Here, the following shorthand notations are used for the matrix elements:

$$\begin{aligned} a_i &\equiv d_i^i \\ b_j^i &\equiv d_j^i - d_{j-1}^i. \end{aligned} \tag{A16}$$

The refractive-index profile can be retrieved by isolating the first term in Eq. (A15) and applying the inverse matrix.

## 6. Overview of Shadowgraphy and Its Limitations

In the main text, shadowgraphy was introduced as a conventional technique for measuring the imaginary component of the refractive index in the vicinity of a laser-induced filament. This section provides a concise overview of this method and discusses its inherent limitations.

Shadowgraphy relies on the quantitative relationship between a material's absorption coefficient and the transmittance of probe light, as described by the Beer–Lambert law.

$$\frac{I_1}{I_0} = \exp\left(-\int \alpha_x \, dx\right), \quad \text{(A17)}$$

where the probe beam is assumed to propagate along the $x$-axis; $\alpha_x$ denotes the absorption coefficient at each $x$-coordinate; and $I_0$ and $I_1$ represent the incident and transmitted light intensities, respectively. Assuming cylindrical symmetry of the filament, the integral in Eq. (A17) can be numerically evaluated by discretizing the axisymmetric distribution of $\alpha$ along the beam path, as detailed in the main text. Under this configuration, the spatial distribution of the imaginary component of the refractive index $k$ can be estimated via the general relation $\alpha = 2\omega k/c$, where $\omega$ is the angular frequency of the probe light and $c$ is the speed of light.

The limitations of shadowgraphy become particularly evident when the probe-light intensity is insufficient. Figure A7(a) shows this effect by presenting the spatial distribution of the imaginary component of the refractive index near a filament induced by a 2-ps USPL under three distinct probe-intensity conditions: high, medium, and low. As shown in the spatial maps, the retrieved profiles are highly sensitive to the probe intensity. For a more quantitative analysis, panel (b) presents the refractive-index values along the filament axis, while panel (c) shows the transverse distributions at the sample interface ($z$ = 0). The results show a consistent trend: lower probe intensities result in a progressive underestimation of the measured refractive-index values.

This effect is the most pronounced in panel (c). Under a sufficiently high probe intensity, the reconstructed radial profile of the imaginary refractive-index component exhibits a Gaussian-like shape, reflecting the spatial intensity profile of the USPL (see the main text). However, as the probe intensity decreases, this correspondence diminishes: the reconstructed values near the filament axis fall below those of the surrounding region, thus resulting in an unphysical distribution. From a physical perspective, the central region of the filament should experience the highest laser intensities and thus the greatest refractive-index modulation. The observed deviation under low probe intensity strongly suggests a measurement artifact, rather than a genuine physical feature.

This discrepancy can be qualitatively explained by the limited precision in determining the transmittance ratio in Eq. (A17) at low probe intensities. Specifically, when the incident intensity $I_0$ becomes small, the signal-to-noise ratio for measuring the transmittance $I_1/I_0$ decreases, thus resulting in increased uncertainty and systematic underestimation in the derived absorption coefficient, and consequently, the imaginary refractive index.

By contrast, the method proposed in this study is intrinsically resistant to such degradation in measurement accuracy arising from variations in probe intensity. This robustness stems from the fact that the probe intensity affects only the amplitude $A$ in Eq. (2), while the refractive index-related parameters $\Psi$ and $\Delta$ remain unaffected. Consequently, the proposed approach enables more reliable extraction of the refractive index, even under suboptimal probe-intensity conditions.

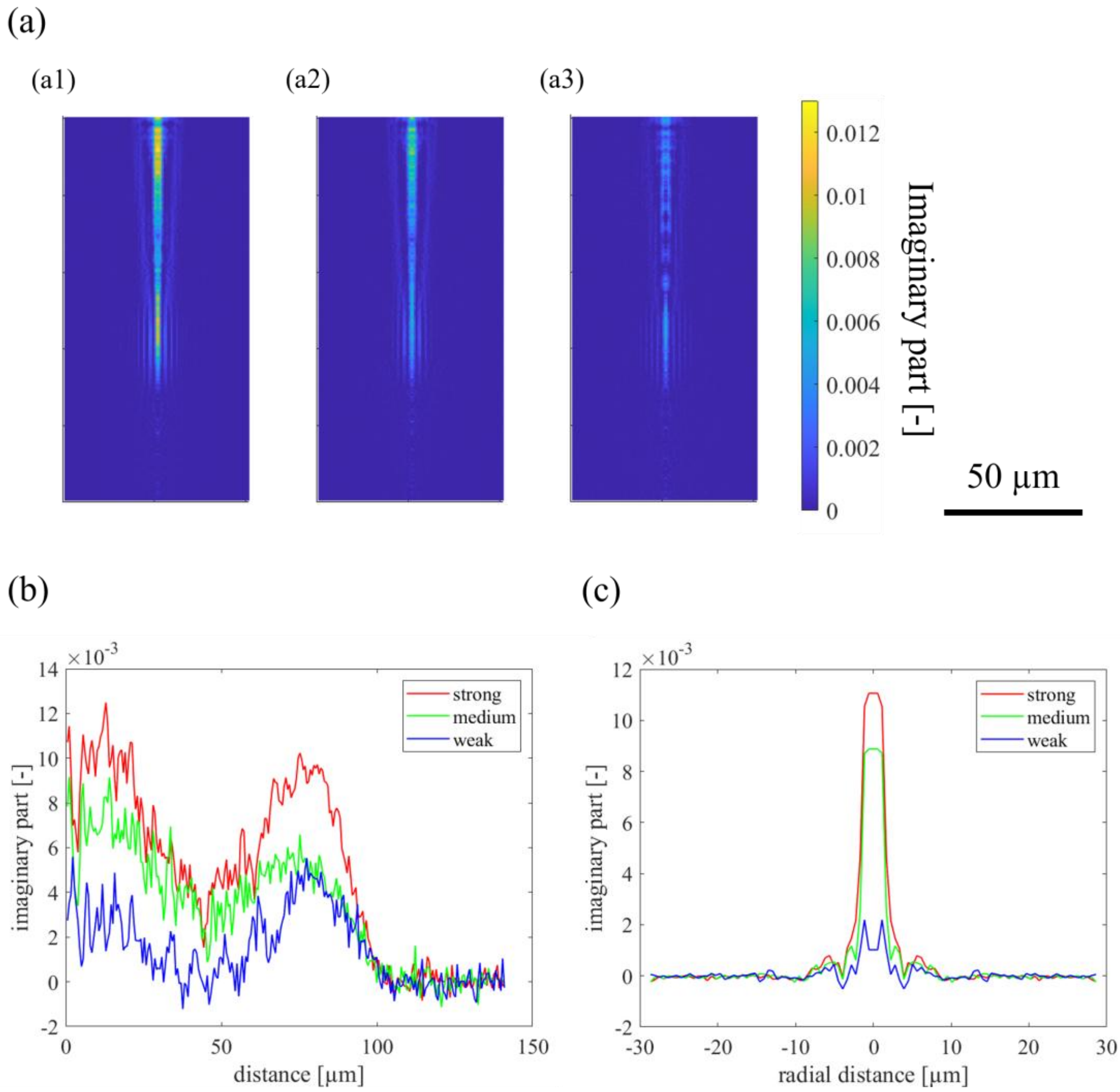

**Fig. A7 Effect of probe intensity on shadowgraphy-based estimation of imaginary refractive index**: (a1–a3) Spatial distributions under high, medium, and low probe intensities, respectively; (b) longitudinal profiles along filament axis; and (c) transverse profiles at sample interface ($z = 0$).

## 7. Overview of Drude Model Used for Real-Component Estimation

This section provides an outline of the Drude model applied to estimate the real component of the refractive index in the main text. This model offers a classical description of the dynamics of photoexcited electrons induced in a dielectric medium. The complex refractive index $N$ is expressed as follows:

$$N = \sqrt{{n_0}^2 - \frac{e^2 n_e}{m_e \varepsilon_0} \frac{1}{\omega^2 + i\omega v}}. \quad \text{(A18)}$$

The parameters in Eq. (A18), along with their physical meanings and typical values, are summarized in Table A2.

**Table A2 Physical parameters used in Drude model.**

| Notation | Description | Typical value |
|---|---|---|
| $N$ | Complex refractive index | N/A |
| $n_e$ | Excited electron density | N/A |
| $v$ | Electron collision frequency | $1 \times 10^{15}$–$1 \times 10^{16}$ [/s] |
| $n_0$ | Refractive index of pristine fused silica | 1.461 [-] |
| $\omega$ | Laser pulsation | $3.66 \times 10^{15}$ [rad/s] |
| $e$ | Elementary charge | $1.602 \times 10^{-19}$ [C] |
| $m_e$ | Electron mass | $9.109 \times 10^{-31}$ [kg] |
| $\varepsilon_0$ | Vacuum permittivity | $8.854 \times 10^{-12}$ [F/m] |

The Drude model is particularly suitable for estimating the real component of the refractive index owing to its characteristic prediction that both the real and imaginary components vary monotonically with the excited electron density. This relationship is illustrated in Fig. A8: Panels (a1) and (a2) show the calculated imaginary and real components, respectively, as a function of $n_e$ at a representative collision frequency of $v = 5 \times 10^{15}$ s$^{-1}$. As $n_e$ increases, the imaginary component increases monotonically while the real component decreases monotonically. This monotonic dependence establishes a one-to-one correspondence under a fixed $v$: a specified imaginary component uniquely determines $n_e$, which in turn determines a unique real component (see the dashed green arrows in panels (a1) and (a2)). This mapping is visualized in panel (a3), and its variation with respect to $v$ is shown in panel (b).

In the main text, this property is leveraged to estimate the real component of the refractive index from experimentally measured values of the imaginary component. The resulting estimates are compared with independent experimental measurements. Since the estimated real component depends on the assumed value of $v$, as discussed previously, agreement between the estimated and measured values—when using $v$ values consistent with prior literature—supports the validity of the proposed experimental method. Nevertheless, the applicability of this method is limited to regimes in which the optical response of the excited region is adequately described by the Drude model. To date, no numerical model accurately captures the carrier dynamics across the full temporal

range—from initial excitation to subsequent relaxation. Therefore, any model-based interpretation must account for these inherent limitations.

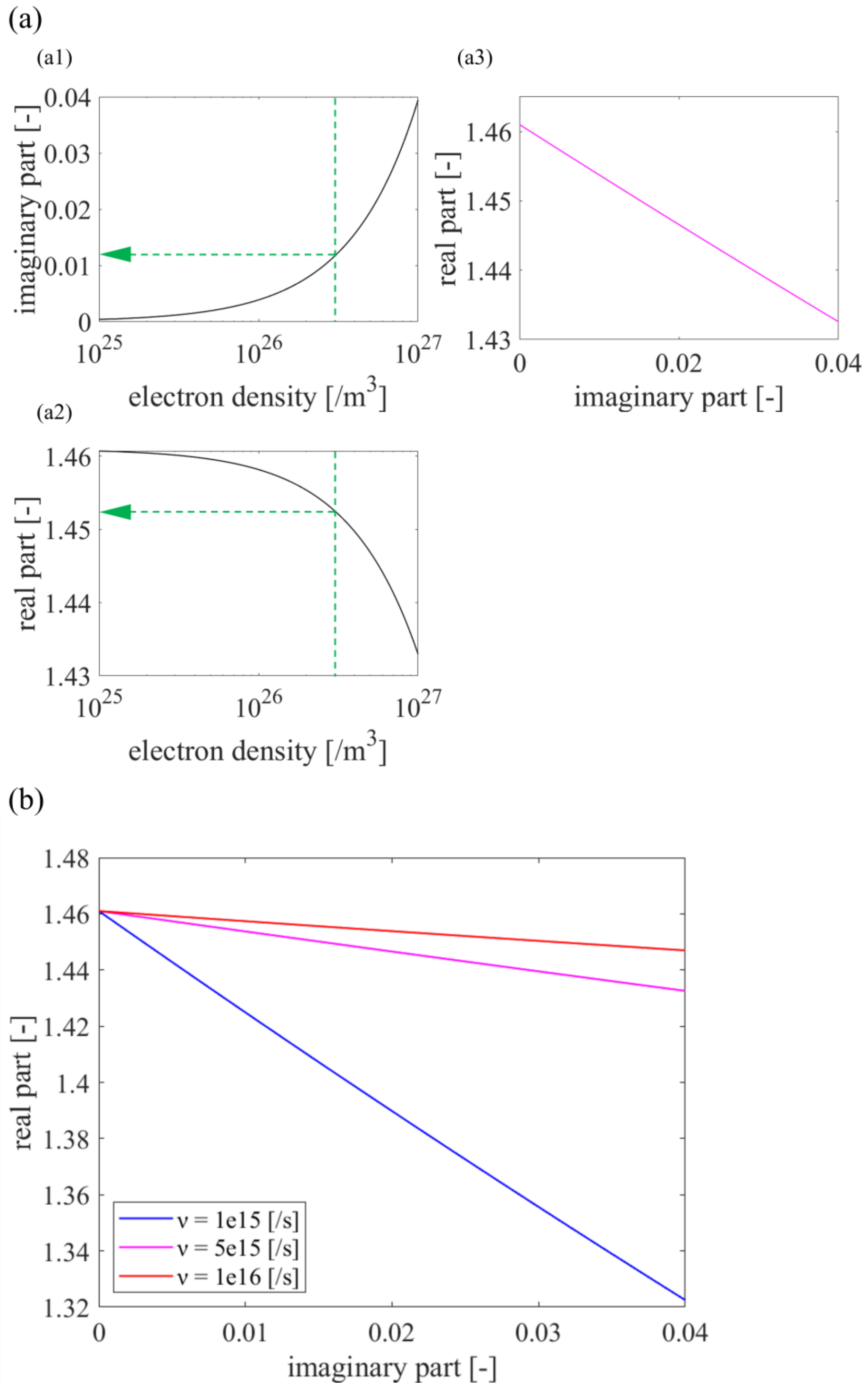

**Fig. A8 Drude Model-based relationship between excited electron density and complex refractive index:** (a1, a2) Imaginary and real components as function of electron density at $v = 5 \times 10^{15}\ s^{-1}$; (a3) one-to-one mapping between imaginary and real components under fixed $v$; (b) dependence of this mapping on collision frequency $v$.